\documentclass[twocolumn,floatfix,prl, aps]{revtex4-2} 
\usepackage{graphicx,amsfonts,amssymb,amsmath,hyperref,hypcap,enumerate,enumitem}
\usepackage{xcolor}
\usepackage{soul}
\usepackage{scalerel}
\usepackage{adjustbox,stmaryrd}
\usepackage{cancel}
\usepackage{csquotes}
\usepackage[utf8]{inputenc}
\usepackage{geometry}
\hypersetup{  colorlinks=true, linkcolor=blue, citecolor=blue, urlcolor=blue  }

\begin{document}
\title{In-plane magnetic field-induced orbital FFLO superconductivity in twisted WSe$_2$ homobilayers}
\author{Jihang Zhu}
\affiliation{Condensed Matter Theory Center and Joint Quantum Institute, Department of Physics, University of Maryland,
College Park, Maryland 20742, USA}
\author{Yang-Zhi Chou}
\affiliation{Condensed Matter Theory Center and Joint Quantum Institute, Department of Physics, University of Maryland,
College Park, Maryland 20742, USA}
\author{Yi Huang}
\affiliation{Condensed Matter Theory Center and Joint Quantum Institute, Department of Physics, University of Maryland,
College Park, Maryland 20742, USA}
\author{Sankar Das Sarma}
\affiliation{Condensed Matter Theory Center and Joint Quantum Institute, Department of Physics, University of Maryland,
College Park, Maryland 20742, USA}

\begin{abstract}
We theoretically predict the in-plane magnetic field-induced orbital Fulde-Ferrell-Larkin-Ovchinnikov (FFLO) superconducting states in twisted WSe$_2$ homobilayers (tWSe$_2$), focusing on its dependence on layer polarization and Fermi surface geometry. 
For unpolarized layers, finite-momentum pairing emerges only at low temperatures and above a critical field $B_{c1,\parallel}$. When layer symmetry is broken, finite-momentum pairing is stabilized at any nonzero field, with a critical temperature higher than that of the zero-momentum state.
Notably, we identify a phase transition, which separates two distinct FFLO phases, when there are two separate Fermi pockets residing in the two moir\'e mini-valleys associated with opposite layers. We further discuss the effects of twist angles and applied field directions.  Our findings establish tWSe$_2$ as a promising platform for realizing and manipulating FFLO states via twist angle, displacement field, and filling factor. 
\end{abstract}

{\let\newpage\relax\maketitle}

\setcounter{page}{1} 

{\em Introduction.}
Recently, robust superconductivity has been observed in twisted WSe$_2$ homobilayers (tWSe$_2$) \cite{CRDean_WSe2_2024, KFMak_WSe2_2024}, with either small or large layer polarization controlled by a displacement field. Notably, superconducting states emerge near highly resistive \cite{CRDean_WSe2_2024} or insulating phases \cite{KFMak_WSe2_2024}, prompting various theoretical studies to understand their origin \cite{Fischer_tWSe2_SC_2024, Schrade_tWSe2_SC_2024, Klebl_tWSe2_SC_2023, Chubukov_tWSe2_SC_2024, Guerci_tWSe2_SC_2024, Qin_tWSe2_SC_2024, JZhu_WSe2_SC_2025, Tuo_tWSe2_SC_2024, Kim_tWSe2_SC_2025, Wietek_tTMD_SC_2022, Christos_tWSe2_SC_2024, Xie_tWSe2_SC_2024}. In transition metal dichalcogenides (TMDs), strong Ising spin-orbit coupling (SOC) locks electron spins to the $\tau z$ direction, where $\tau=\pm 1$ encodes the valley. 
An in-plane magnetic field $B_\parallel$, albeit much weaker than the Ising SOC effect ($\sim 500$ meV \cite{Le_WSe2_SOC_2015} in tWSe$_2$), shifts the layer-dependent canonical momenta and deforms the Fermi surface (FS) through interlayer coupling. Such $B_\parallel$-induced orbital effects favor finite-momentum Cooper pairing and provide a new pathway for realizing Fulde–Ferrell–Larkin–Ovchinnikov (FFLO) superconductivity \cite{FF_1964, LO_1965} purely through orbital effects \cite{PWan_obitalFFLO_2023,CXLiu_FFLO_orb_2017, YMXie_FFLO_2023, Yuan_orbitalFFLO_2023, Yuan_orbitalFFLO_2025, Zhao_orbitalFFLO_2023, Cao_orbitalFFLO_2024, Zhao_orbitalFFLO_2024, Cho_orbitalFFLO_2024, Yang_orbitalFFLO_2024, Qiu_orbitalFFLO_2022, Yan_orbitalFFLO_2024, Nag_orbitalFFLO_2024}.

The FFLO state \cite{FF_1964, LO_1965} was originally proposed for $s$-wave spin-singlet superconductors under a Zeeman field,
where spin-split Fermi surfaces (FSs) favor finite-momentum pairing at low temperatures and high magnetic fields exceeding the Pauli paramagnetic limit \cite{Chandrasekhar_1962, Clogston_1962}.
While most studies have focused on Zeeman-driven spin mechanisms \cite{Agterberg_PDW_2020, Casalbuoni_FFLO_2004, Radzihovsky_2010, Houzet_FFLO2D_2002,Klemm_Hc2_1975, Klemm_Hc2_1974, Barzykin_FFLO_SOC_2002, Zheng_FFLO_SOC_2014, NYuan_FFLO_SOC_2021}, recent work on Ising superconductors has revealed FFLO phenomena arising from both spin and orbital effects \cite{XxXi_NbSe2_2016,Saito_Ising_2016, Lu_Ising_2015, Sohn_Ising_2018, Barrera_Ising_2018, YMWu_PDW_2023, YMWu_PDWvHs_2023}.
The purely orbital FFLO mechanism in tWSe$_2$ under $B_\parallel$ \cite{YMXie_FFLO_2023} has several distinct features: (i) its two-dimensional (2D) nature mitigates vortex formation, (ii) the system is in the clean limit, with a mean free path much longer than the superconducting coherence length \cite{CRDean_WSe2_2024}, and (iii) the FS is highly tunable via displacement field and doping, allowing rich phenomena driven by layer polarization and FS geometry \cite{Zhang_TMDhomo_2024, FWu_TMDhomoBi_2019, TDevakul_DFT_2021, HYu_TMDhomo_2019, DZhai_TMDhomo_2020, Naik_TMDhomo_2018, Kundu_TMDhomo_2022}.

\begin{figure}[!t]
\centering
\includegraphics[width=1.0\columnwidth]{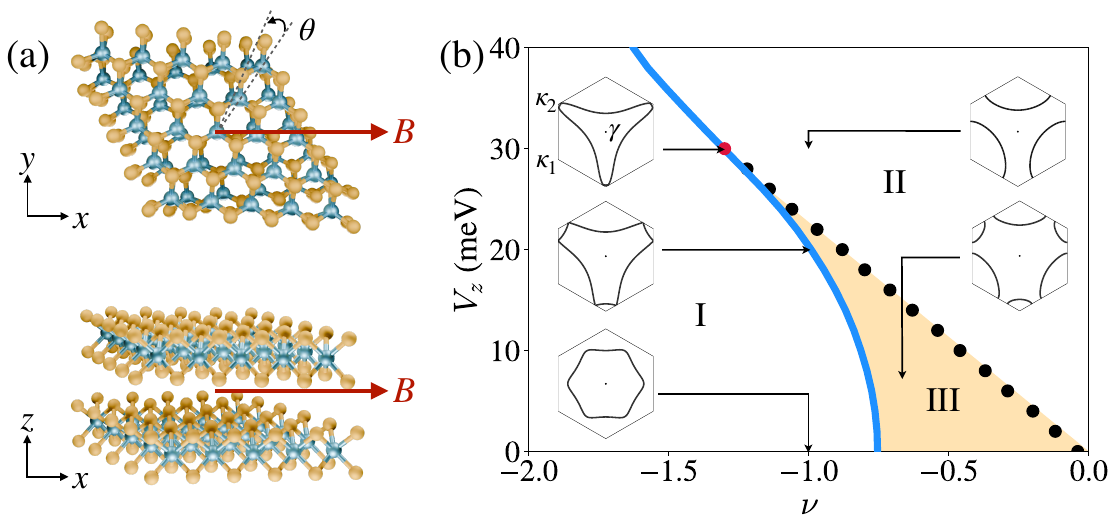}
\caption{\label{fig1_phasediagram} {
(a) Top and side views of tWSe$_2$, with tungsten (blue) and selenium (gold) atoms.
(b) Phase diagram of $3.65^\circ$ tWSe$_2$ as a function of electron filling factor $\nu$ (counting two spin-valley flavors) and displacement field (characterized by the interlayer energy difference $V_z$), showing different FS geometries. 
Region I: A single FS at $\gamma$. 
Region II: A single FS at $\kappa_1$ ($V_z > 0$). 
Region III (shaded yellow): Two Fermi pockets at $\kappa_1$ and $\kappa_2$, where a phase transition occurs under increasing $B_\parallel$, separating two distinct FF phases, but only when $V_z \neq 0$.
The blue line marks the van Hove singularity (VHS). Black dots mark numerically determined boundaries of region III, terminating at a higher-order VHS (red dot) \cite{Hsu_VHS_2021, YMWu_PDWvHs_2023}. 
}}
\end{figure}

In this Letter, we systematically investigate the orbital FFLO superconducting state induced by $B_\parallel$ in tWSe$_2$, as illustrated in Fig.~\ref{fig1_phasediagram}(a).
We find that the upper critical field $B_{c2,\parallel}$ and the nature of the FFLO phase depend sensitively on layer polarization and FS geometry. 
For unpolarized layers, finite-momentum pairing emerges only at $T/T_{c0} < \alpha$ (where $\alpha<1$ and $T_{c0}$ is the zero-field $T_c$) and above a critical field $B_{c1,\parallel}$.
In contrast, when layer symmetry is broken, finite-momentum pairing can be stabilized at any $T/T_{c0} < 1$ and nonzero $B_{\parallel}$. 
Remarkably, when two separate Fermi pockets reside in two moir\'e mini-valleys [region III in Fig.~\ref{fig1_phasediagram}(b)], a phase transition separating two distinct Fulde–Ferrell (FF) phases occurs. 
Our findings highlight moir\'e Ising superconductors as a unique platform for realizing and manipulating FFLO states via displacement field, filling factor and twist angle.







{\em Twisted WSe$_2$ under an in-plane magnetic field.}
Due to strong Ising SOC, spin-valley locking in tWSe$_2$ suppresses the Zeeman effect from $B_\parallel$, which we neglect.
Under an applied field $\pmb{B} = B_x \hat{x} + B_y \hat{y}$, the system is described by the Hamiltonian
\begin{equation}
\label{Eq_Hamil}
\setlength\arraycolsep{-8pt} 
\begin{split}
\mathcal{H} = 
\begin{bmatrix}
h_{\pmb{q}_1 + \delta\pmb{k}_1}
+ U_1(\pmb{r})
& T(\pmb{r}) \\
T^\dagger(\pmb{r}) & 
h_{\pmb{q}_2 + \delta\pmb{k}_2}
+ U_2(\pmb{r})
\end{bmatrix} 
+ \frac{V_z}{2} \hat{l}_z.
\end{split}
\end{equation}
Focusing on hole doping, $h_{\pmb{q}} = -\hbar^2q^2/2m^*$, $\pmb{q}_l = \pmb{k}-\pmb{\kappa}_l$ is the momentum relative to the Dirac points $\pmb{\kappa}_l$ of layer $l=1,2$. The interlayer energy difference $V_z$ arises from the applied displacement field, and $\hat{l}_z$ is the Pauli matrix acting on the layer degree of freedom.
The moir\'e potentials $U_l(\pmb{r})$ and interlayer tunneling $T(\pmb{r})$ take the forms 
$U_{1,2}(\pmb{r}) = 2V\sum\limits_{j=1,3,5} \cos(\pmb{G}_j \cdot \pmb{r} \mp \psi)$, $T(\pmb{r}) = w(1+e^{-i\pmb{G}_2 \cdot \pmb{r}}+e^{-i\pmb{G}_3 \cdot \pmb{r}})$ \cite{continuum_paras, TDevakul_DFT_2021, JZhu_WSe2_SC_2025}.
We focus on the twist angle $\theta=3.65^\circ$ and general properties of orbital FFLO states in this Letter, results of other twist angles can be found in SM~\cite{SeeSM}.
To leading order, $T(\pmb{r})$ is local and unaffected by $B_\parallel$, which induces a momentum shift $\delta \pmb{k}_l = e\pmb{A}_l/\hbar$ in layer $l$ \cite{FWu_TBG_Bpara_2019, WQin_TTG_Bpara_2021, ELake_TTG_SC_2021} via the Peierls phase. In the Landau gauge, $\pmb{A}_l = B_y z_l \hat{x}-B_xz_l \hat{y}$.
Since the two layers are separated along the out-of-plane ($z$) direction, the momentum shifts satisfy
\begin{equation}
\begin{split}
\delta\pmb{k}_1 - \delta\pmb{k}_2
&= q_B (\sin\theta_B \hat{x} - \cos\theta_B\hat{y}),
\end{split}
\end{equation}
where $\theta_B$ is the angle of $\pmb{B}$ relative to the $x$-axis, and
$q_B = edB/\hbar = d/l_B^2$, with magnetic length $l_B = \sqrt{\hbar/eB}$ and layer separation $d=0.7$ nm in tWSe$_2$.
Due to gauge freedom, in our calculations we choose
\begin{align}
\label{Eq_q1q2}
\delta\pmb{k}_1 &= \frac{q_B}{2} \Big[ (1-\bar{l}_z) \sin\theta_B  \hat{x}-(1-\bar{l}_z) \cos\theta_B \hat{y} \Big],\\
\delta\pmb{k}_2 &= \frac{q_B}{2} \Big[-(1+\bar{l}_z) \sin\theta_B \hat{x} + (1+\bar{l}_z) \cos\theta_B  \hat{y} \Big]. \nonumber
\end{align}
$\bar{l}_z \in [-1,1]$ is the average layer polarization, defined as
$\bar{l}_z = \sum\limits_{n \pmb{k}} \langle n \pmb{k}|\hat{l}_z |n\pmb{k} \rangle f_{n\pmb{k}}/ \sum\limits_{n \pmb{k}} f_{n\pmb{k}}$ for electron doping and 
$\bar{l}_z = \sum\limits_{n \pmb{k}} \langle n \pmb{k}|\hat{l}_z |n\pmb{k} \rangle (1-f_{n\pmb{k}})/ \sum\limits_{n \pmb{k}} (1-f_{n\pmb{k}})$ for hole doping.
By choosing Eq.~(\ref{Eq_q1q2}), we ensure that $B_\parallel$ does not influence the superconducting state through orbital effects when carriers are fully layer-polarized.


{\em Finite-momentum pairing.}
We focus on the intervalley intralayer phenomenological BCS pairing interaction within the continuum model framework \cite{JZhu_WSe2_SC_2025}. Specifically, we assume that $B_\parallel$ does not change the coupling constant of the pairing interaction.
Other cases will be discussed at the end of this Letter.

We consider an intervalley intralayer pairing with local attraction \cite{JZhu_WSe2_SC_2025}.
The interacting Hamiltonian for a Cooper pair with momentum $\pmb{q}$ is
\begin{align}
\mathcal{H}_{\rm int} = -\frac{g}{A} &\sum\limits_{\substack{l,\pmb{k},\pmb{k}'\\\pmb{G},\pmb{G}'}}
\psi^{\dagger}_{+,\pmb{G},l}(\pmb{k}+\frac{\pmb{q}}{2}) \psi^{\dagger}_{-,-\pmb{G},l}(-\pmb{k}+\frac{\pmb{q}}{2}) \nonumber \\
&\psi_{-,-\pmb{G}',l}(-\pmb{k}'+\frac{\pmb{q}}{2}) \psi_{+,\pmb{G}',l}(\pmb{k}'+\frac{\pmb{q}}{2}),
\end{align}
where $\psi^\dagger_{\tau,l}$ is the creation operator for electrons in valley $\tau = \pm$ and layer $l$, $A$ is the sample area, $\pmb{G}$ and $\pmb{G}'$ are moir\'e reciprocal lattice vectors. 
In this Letter, we use $g=120$meV$\cdot$nm$^2$ \cite{JZhu_WSe2_SC_2025} for our calculations. The qualitative results, expressed in dimensionless units, do not depend on the specific choice of $g$.
Using the single-band approximation, the band-projected interaction takes the form
\begin{align}
\mathcal{H}_{\rm int} = -\frac{g}{A} \sum\limits_{l,\pmb{k},\pmb{k}'} u^*_{\pmb{k},l}&(\pmb{q})u_{\pmb{k}',l}(\pmb{q})
c^\dagger_+(\pmb{k}+\frac{\pmb{q}}{2}) c^\dagger_-(-\pmb{k}+\frac{\pmb{q}}{2}) \nonumber\\
&c_-(-\pmb{k}'+\frac{\pmb{q}}{2}) c_+(\pmb{k}'+\frac{\pmb{q}}{2}),
\end{align}
where $c^\dagger_{\tau}(\pmb{k}) = \sum\limits_{l,\pmb{G}} z_{\tau\pmb{G}l}(\pmb{k}) \psi^\dagger_{\tau\pmb{G}l}(\pmb{k})$ is the creation operator in band representation,
and $u_{\pmb{k},l}(\pmb{q}) = \sum\limits_{\pmb{G}} z_{+,\pmb{G},l}(\pmb{k}+\pmb{q}/2) z_{-,-\pmb{G},l}(-\pmb{k}+\pmb{q}/2)$.

\begin{figure}[!b]
\centering
\includegraphics[width=1.0\columnwidth]{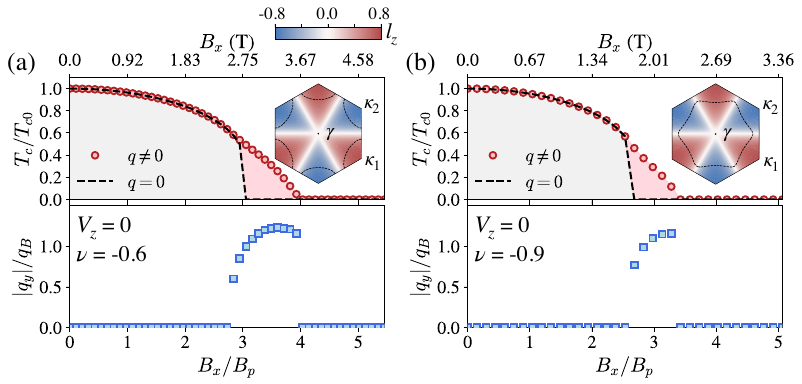}
\caption{\label{fig_TcvsB_Vz0} {
$T_c$ vs. $B_x$ and the Cooper pair momentum corresponding to the highest $T_c$ of $3.65^\circ$ tWSe$_2$ for layer symmetric case $V_z=0$ at
(a) $\nu = -0.6$ and (b) $\nu = -0.9$.
Hexagonal insets: layer polarization $l_z(\pmb{k})$ in the moir\'e Brillouin zone (MBZ), sharing the same colorbar on the top left. Black dashed contours depict corresponding FSs.
The top $x$-axes show $B_x$ in Tesla for reference, using an attraction strength $g=120$ meVnm$^2$.
}}
\end{figure}

The Bogoliubov–de Gennes Hamiltonian with spinor $\Psi_{\pmb{k}}^{\dagger} = (c^\dagger_+(\pmb{k}+\pmb{q}/2) \quad c_-(-\pmb{k}+\pmb{q}/2))$ is
\begin{equation}
H_{\rm BdG}({\pmb{k}}) = 
\begin{pmatrix}
\xi_{+}(\pmb{k}+\frac{\pmb{q}}{2}) & \Delta_{\pmb{k},\pmb{q}} \\
\Delta^*_{\pmb{k},\pmb{q}} & -\xi_{-}(-\pmb{k}+\frac{\pmb{q}}{2})
\end{pmatrix},
\end{equation}
$\xi_{\tau} = \varepsilon_{\tau} - \mu$ is the band energy measured from the chemical potential. $\Delta_{\pmb{k},\pmb{q}} = \sum\limits_{l} u^*_{\pmb{k},l}(\pmb{q}) \tilde{\Delta}_{\pmb{q},l}$ is the superconducting order parameter.
We define the layer-resolved superconducting order parameter
\begin{equation}
\begin{split}
\tilde{\Delta}_{\pmb{q},l} 
&= -\frac{g}{A}\sum\limits_{\pmb{k}} u_{\pmb{k},l}(\pmb{q}) \langle c_-(-\pmb{k}+\frac{\pmb{q}}{2}) c_+(\pmb{k}+\frac{\pmb{q}}{2}) \rangle.
\end{split}
\end{equation}
The linearized gap equation is
\begin{equation}
\tilde{\Delta}_{\pmb{q},l} = \sum\limits_{l'} \chi_{ll'}(\pmb{q}) \tilde{\Delta}_{\pmb{q},l'},
\end{equation}
where $\chi_{ll'}$ is the matrix element of the superconducting pair susceptibility
\begin{align}
\label{Eq_chillp}
\chi_{ll'}(\pmb{q}) = &\frac{g}{A} \sum\limits_{\pmb{k}'} 
u_{\pmb{k}',l} (\pmb{q},\pmb{B}) u^*_{\pmb{k}',l'} (\pmb{q},\pmb{B}) \\
&\frac{1-f(\xi_+(\pmb{k}'+\frac{\pmb{q}}{2},\pmb{B}))-f(\xi_-(-\pmb{k}'+\frac{\pmb{q}}{2},\pmb{B}))}{\xi_+(\pmb{k}'+\frac{\pmb{q}}{2},\pmb{B}) + \xi_-(-\pmb{k}'+\frac{\pmb{q}}{2},\pmb{B})}, \nonumber
\end{align}
with Fermi-Dirac distribution $f(\xi) = [e^{\xi/k_BT}+1]^{-1}$.
The superconducting transition temperature $T_c = \text{max}\{T_c(\pmb{q})\}$ \cite{Holleis_BG_SC_2025}, and $T_c(\pmb{q})$ is the temperature at which the largest eigenvalue of matrix $\pmb{\chi}$ reaches one.
To streamline numerical calculations, we approximate $\varepsilon(\pmb{k},\pmb{B})$ and $u_{\pmb{k},l}(\pmb{q}, \pmb{B})$ using small-momentum and weak-field expansions \cite{SeeSM}:
\begin{align}
\label{Eq_expansion}
u_{\pmb{k},l}(\pmb{q}, \pmb{B}) &\approx  \sum\limits_{\pmb{G}} |z_{+,l,\pmb{G}}(\pmb{k})|^2, \\
\varepsilon(\pmb{k}_0 + \pmb{q}, \pmb{B}) &\approx \varepsilon(\pmb{k}_0,B=0) + \pmb{v}(\pmb{k}_0) \cdot \pmb{q} + \pmb{M}(\pmb{k}_0) \cdot \pmb{B}, \nonumber
\end{align}
where $\pmb{v}(\pmb{k}) = \partial_{\pmb{k}} \varepsilon|_{B=0}$, $\pmb{M}(\pmb{k}) = \partial_{\pmb{B}} \varepsilon|_{B=0}$, which are provided in the SM \cite{SeeSM} for twist angles $2^\circ, 3.65^\circ$ and $6^\circ$. 
We have numerically checked several cases using full calculations without performing expansions.

{\em Layer symmetric case }($V_z=0$).
We first discuss the zero displacement field case, where the system remains layer-unpolarized with $\bar{l}_z = 0$ at any filling.
Figure~\ref{fig_TcvsB_Vz0} shows $T_c$ as a function of $\pmb{B} = B_x \hat{x}$ \footnote{Due to the $C_{3z}$ rotational symmetry of tWSe$_2$, the critical in-plane field remains the same for equivalent directions.} for two different filling factors, $\nu=-0.6, -0.9$, with distinct FSs as shown in the insets of Fig.~\ref{fig_TcvsB_Vz0}.
Despite the differences in the FS geometry at these fillings, the phase diagrams remain qualitatively similar due to the preserved layer symmetry.
In particular, phase diagrams in Fig.~\ref{fig_TcvsB_Vz0} are reminiscent of that in Zeeman-driven FFLO systems \cite{Casalbuoni_FFLO_2004, Shimahara_FF_1994,Radzihovsky_2010,Agterberg_PDW_2020, Radzihovsky_LO_2011, KYang_FFLO_2001}.
At small $B_\parallel$, the system remains in a BCS-type ($q=0$) superconducting phase (gray region in Fig.~\ref{fig_TcvsB_Vz0}). However, for temperatures below $T/T_{c0} \sim 0.56$ and high fields $B_x/B_p \gtrsim 3$ \footnote{The tricritical point temperature of $3.65^\circ$ tWSe$_2$, $T_{\rm tric}/T_{c0} \approx 0.56$, coincidentally matches that of Ref.\cite{Casalbuoni_FFLO_2004}. However, in tWSe$_2$, this value depends on the interlayer coupling strength: as the interlayer coupling decreases (or the twist angle increases), the tricritical points shifts to a higher
$T_{\rm tric}/T_{c0}$ and lower $B/B_p$.
Unlike in conventional Zeeman-driven FFLO phase diagram, where the transition from $q=0$ to $q\neq 0$ superconductivity at $T=0$ occurs at the Pauli limit $B/B_p = 1$, in Ising superconductors, this transition takes place at a higher $B/B_p$ due to orbital effects and Ising SOC.}, the system transitions into a finite-$q$ superconducting state (pink region in Fig.~\ref{fig_TcvsB_Vz0}), where Cooper pairs acquire a finite momentum $q \sim q_B = d/l^2_B$ which is perpendicular to the applied in-plane field (lower panels of Fig.~\ref{fig_TcvsB_Vz0}).
Here, $T_{c0}$ is the zero-field critical temperature, and $B_p = 1.86 T_{c0}$ denotes the Pauli paramagnetic limit.
For the layer-unpolarized case, the superconducting order parameters associated with opposite momenta $\Delta_{\pm \pmb{q}}$ are nearly degenerate \cite{YMXie_FFLO_2023}, so only $|q_y|$ is shown in Fig.~\ref{fig_TcvsB_Vz0}. As discussed in Ref.~\cite{YMXie_FFLO_2023}, it is believed that FF state is more energetically favorable than the Larkin-Ovchinnikov state in tWSe$_2$.

\begin{figure}[!t]
\centering
\includegraphics[width=1.0\columnwidth]{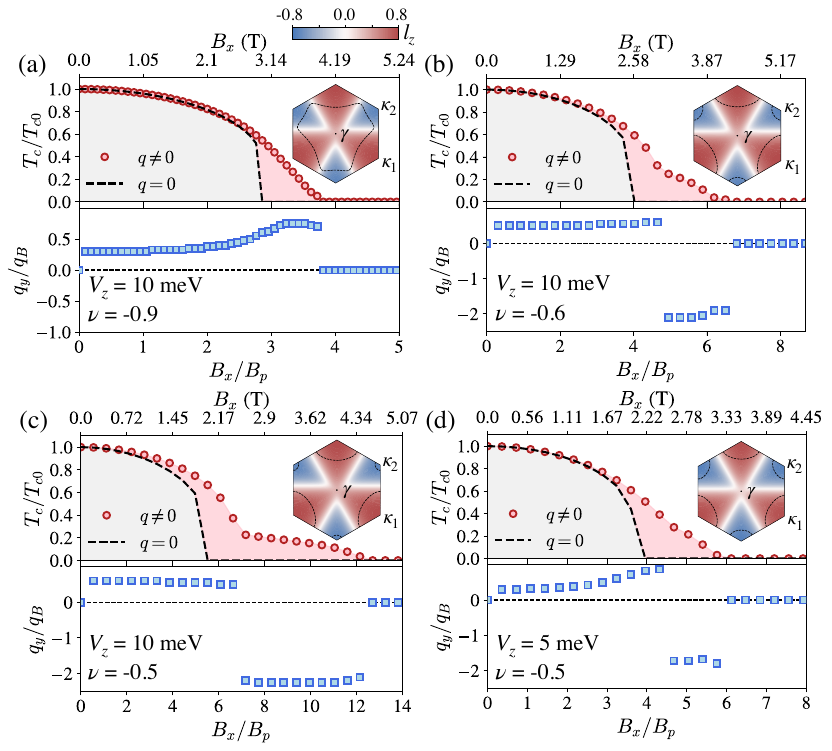}
\caption{\label{fig_qjump} {
$T_c$ vs. $B_x$ and the Cooper pair momentum corresponding to the highest $T_c$ of $3.65^\circ$ tWSe$_2$ for layer asymmetric cases. $V_z$ and $\nu$ are indicated in each subfigure.
An FF phase can be stabilized at any nonzero $B_x$. 
The evolution of Cooper pair momentum with $B_x$ depends on the FS geometry.
}}
\end{figure}

\begin{figure}[!t]
\centering
\includegraphics[width=1.0\columnwidth]{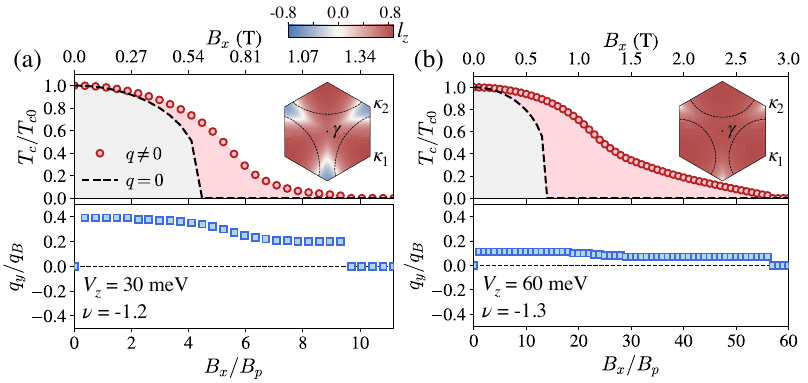}
\caption{\label{fig_TcvsB_Vz30to60} {
$T_c$ vs. $B_x$ and the Cooper pair momentum corresponding to the highest $T_c$ in $3.65^\circ$ tWSe$_2$ for large displacement fields.
These filling factors are chosen to maintain similar FS geometries.}
}
\end{figure}

{\em Layer asymmetric case} ($V_z>0$).
A finite displacement field breaks layer symmetry, causing carriers to favor one layer (layer $l=1$ for $V_z>0$). Unlike the layer symmetric case [Fig.~\ref{fig_TcvsB_Vz0}], where the FF phase requires a finite $B_\parallel$, finite-$q$ pairing is stabilized at any nonzero $B_\parallel$, as shown in Figs.~\ref{fig_qjump}-\ref{fig_TcvsB_Vz30to60}. Notably, the degeneracy between $\pmb{q}$ and $-\pmb{q}$ is lifted, and all finite-$q$ pairing states are FF states.

For a small displacement field, $V_z \lesssim 20$ meV, the Cooper pair momentum evolves with increasing $B_x$, depending on the FS geometry.
When two separate FSs exist at $\kappa_1$ and $\kappa_2$, associated with opposite layers [Fig.~\ref{fig_qjump}(b-d)], the phase diagram features a transition between two distinct FF phases with different pairing momenta. This phase transition is marked by a discontinuous jump in $\pmb{q}$, particularly in its direction. 
At small $B_x$, pairing near $T_c$ is dominated by carriers from $\kappa_1$, where larger layer polarization results in a smaller $q$. As $B_x$ increases, the dominant pairing shifts to $\kappa_2$, which has a larger $q$.
This phase transition is more pronounced when the two FSs differ significantly in layer polarization [Fig.~\ref{fig_qjump}(b-d)], and its location in $(\nu, V_z)$ phase diagram is highlighted as yellow-shaded region in Fig.~\ref{fig1_phasediagram}(b). 
In contrast, when a single FS is present [Fig.~\ref{fig_qjump}(a)], the Cooper pair momentum evolves smoothly with $B_x$ within the FF phase, reflecting the gradual change in the dominant layer contribution to pairing.

As layer polarization increases with larger $V_z$, the FF phase expands over a broader region of the phase diagram, as shown in Fig.~\ref{fig_TcvsB_Vz30to60}. 
At low temperatures, both the critical field for the transition from $q=0$ to $q\neq0$ superconducting phase ($B_{c1,\parallel}$, gray to pink) and the transition from $q\neq0$ superconducting phase to the normal state ($B_{c2,\parallel}$, pink to white) increase with $V_z$. Remarkably, $B_{c2,\parallel}$ can be several times higher than $B_{c1,\parallel}$, indicating a significantly extended FF phase.
However, the increasing layer polarization leads to a reduction in the Cooper pair momentum $q$, this trend is illustrated in Fig.~\ref{fig_TcvsB_Vz30to60} for $V_z=30$ to $60$ meV.

{\em Discussion.} 
The most interesting prediction in this Letter is the emergence of a phase transition within the FF phase, characterized by a sign-changing jump in $\pmb{q}$. 
This transition occurs when the layer symmetry is broken and two distinct FSs associated with different layers are present [Region III in Fig.~\ref{fig1_phasediagram}(b)].
Although $T_c$ does not show discontinuity at this transition, the superconducting diode effect can probe the change of $\pmb{q}$ by measuring supercurrents in two opposite directions (typically perpendicular to the applied magnetic field) \cite{Ando_diode_2020, Bauriedl_diode_2022, Baumgartner_diode_2022, Lin_diode_2022, Pal_diode_2022, YMXie_FFLO_2023, NYuan_diode_2022, Nakamura_diode_2024, He_diode_2022, Bankier_diode_2025}.

Next, we discuss the dependence on the twist angle $\theta$. For smaller twist angles ($\theta \lesssim 4^\circ$), where the moir\'e bands are flatter, an in-plane field has relatively stronger orbital effects, distorting the FS more significantly and resulting in a smaller region for FF states. 
As $\theta$ increases, the FF phase extends over a broader region in the $(B,T)$ phase diagram, with the tricritical point at $V_z=0$ moving to higher temperatures and smaller $B$ fields, suggesting a smaller in-plane magnetic field to induce FF state. The phase diagrams for different $\theta$ can be found in SM \cite{SeeSM}. A related situation is filling near the VHS, where the dispersion becomes significantly flat. In such cases, the FF phase is often suppressed \cite{SeeSM}.

The direction of the applied in-plane magnetic field also plays an important role in shaping the FF phase. 
The orbital effects of $B_x$ and $B_y$ result in different FS deformations (see Figs.~\ref{figS1_Mv}-\ref{figS2_FS} in SM \cite{SeeSM}). A clear example occurs when a single Fermi pocket is around the $\gamma$ point at $V_z=0$: Under $B_x$, the FS undergoes distortions that create nearly ``nesting'' Fermi sheets separated vertically, favoring a finite-$q_y$ pairing; In contrast, the FS reshaping under $B_y$ is very different, making the FF phase less favorable in this case. Further details on the effects of $B_y$, including the corresponding phase diagrams, are provided in the SM \cite{SeeSM}. For $V_z \neq 0$, the dependence on field direction becomes more complex, influenced by detailed features of the FS, layer polarization, and $\pmb{v}(\pmb{k})$ and $\pmb{M}(\pmb{k})$ distributions.
Nevertheless, the general properties of the FF phase we described in Figs.~\ref{fig_TcvsB_Vz0}-\ref{fig_TcvsB_Vz30to60}, including the phase transition within the FF phase, remain valid for $\pmb{B} = B_y \hat{y}$.

An implicit assumption in this Letter is that the pairing strength is \textit{not} affected by $B_\parallel$. Several recent studies suggest the possibility of spin-fluctuation mechanism \cite{Chubukov_tWSe2_SC_2024,Fischer_tWSe2_SC_2024} (as inferred by the moir\'e Hubbard model~\cite{PanH2020}), and Ref.\cite{CRDean_WSe2_2024} reports a nearby antiferromagnetic state with a critical temperature of $\sim 20$K, indicating that spin fluctuations are likely active in the parameter regime we discuss in this work. This supports our assumption that the pairing strength remains independent of $B_\parallel$. A potential caveat is that the antiferromagnetic order could substantially reconstruct the Fermi surface, whereas our theory here is based on the single-particle band structure. 
Furthermore, although different theories \cite{Fischer_tWSe2_SC_2024, JZhu_WSe2_SC_2025} propose different pairing symmetries, the orbital FFLO states captured by a finite-$q$ pairing with enhanced $T_c$ should be generic near the transition point, not tied to a specific pairing symmetry.
A systematic study incorporating different pairing mechanisms, symmetries, and reconstructed FS is an important future direction, but our predicted FFLO physics in an in-plane field should be generic and robust.

Finally, we comment on the $B_\parallel$ response for 2D superconductivity in graphene-based materials. Graphene multilayers present additional complications due to significant Zeeman effects.
An intriguing exception is the recently discovered intravalley spin-polarized superconductivity in electron-doped rhombohedral graphene tetralayers and pentalayers \cite{LongJu_chiralSC}, which has motivated a number of theoretical studies \cite{YChou_chiralSC_2024, Yoon_QM_SC_2025, Kim_chiralSC_2024, Qin_chiralSC_2025, HYang_FFLO_2024, Jahin_chiralSC_2024, Sau_vortex_2024, Wang_chiralSC_2024, Geier_chiralSC_2024, Guinea_chiralSC_2025, MayMann_chiralSC_2025, Gaggioli_pentaG_2025,Dong2025}.
In particular, incommensurate finite-$\pmb{q}$ pairing is expected even without $B_\parallel$ \cite{HYang_FFLO_2024, Christos_pentaG_2025, Chen_pentaG_2025}. Experiments have reported in-plane critical fields far exceeding the Pauli limit, suggesting a rich phase diagram for orbital FF states. The superconducting diode effect \cite{Ando_diode_2020, Bauriedl_diode_2022, Baumgartner_diode_2022, Lin_diode_2022, Pal_diode_2022, YMXie_FFLO_2023, NYuan_diode_2022, Nakamura_diode_2024, He_diode_2022, Bankier_diode_2025} provides a promising avenue to probe these states \cite{Sedov_probeFFLO_2025}.

    {\em Acknowledgments.}
We thank Cory Dean, Liang Fu, Jay D. Sau, Puhua Wan, and Yinjie Guo for valuable discussions. This work is supported by the Laboratory for Physical Sciences.


\begin{thebibliography}{94}%
	\makeatletter
	\providecommand \@ifxundefined [1]{%
		\@ifx{#1\undefined}
	}%
	\providecommand \@ifnum [1]{%
		\ifnum #1\expandafter \@firstoftwo
		\else \expandafter \@secondoftwo
		\fi
	}%
	\providecommand \@ifx [1]{%
		\ifx #1\expandafter \@firstoftwo
		\else \expandafter \@secondoftwo
		\fi
	}%
	\providecommand \natexlab [1]{#1}%
	\providecommand \enquote  [1]{``#1''}%
	\providecommand \bibnamefont  [1]{#1}%
	\providecommand \bibfnamefont [1]{#1}%
	\providecommand \citenamefont [1]{#1}%
	\providecommand \href@noop [0]{\@secondoftwo}%
	\providecommand \href [0]{\begingroup \@sanitize@url \@href}%
	\providecommand \@href[1]{\@@startlink{#1}\@@href}%
	\providecommand \@@href[1]{\endgroup#1\@@endlink}%
	\providecommand \@sanitize@url [0]{\catcode `\\12\catcode `\$12\catcode
		`\&12\catcode `\#12\catcode `\^12\catcode `\_12\catcode `\%12\relax}%
	\providecommand \@@startlink[1]{}%
	\providecommand \@@endlink[0]{}%
	\providecommand \url  [0]{\begingroup\@sanitize@url \@url }%
	\providecommand \@url [1]{\endgroup\@href {#1}{\urlprefix }}%
	\providecommand \urlprefix  [0]{URL }%
	\providecommand \Eprint [0]{\href }%
	\providecommand \doibase [0]{http://dx.doi.org/}%
	\providecommand \selectlanguage [0]{\@gobble}%
	\providecommand \bibinfo  [0]{\@secondoftwo}%
	\providecommand \bibfield  [0]{\@secondoftwo}%
	\providecommand \translation [1]{[#1]}%
	\providecommand \BibitemOpen [0]{}%
	\providecommand \bibitemStop [0]{}%
	\providecommand \bibitemNoStop [0]{.\EOS\space}%
	\providecommand \EOS [0]{\spacefactor3000\relax}%
	\providecommand \BibitemShut  [1]{\csname bibitem#1\endcsname}%
	\let\auto@bib@innerbib\@empty
	\bibitem [{\citenamefont {Guo}\ \emph {et~al.}(2025)\citenamefont {Guo},
		\citenamefont {Pack}, \citenamefont {Swann}, \citenamefont {Holtzman},
		\citenamefont {Cothrine}, \citenamefont {Watanabe}, \citenamefont
		{Taniguchi}, \citenamefont {Mandrus}, \citenamefont {Barmak}, \citenamefont
		{Hone}, \citenamefont {Millis}, \citenamefont {Pasupathy},\ and\
		\citenamefont {Dean}}]{CRDean_WSe2_2024}%
	\BibitemOpen
	\bibfield  {author} {\bibinfo {author} {\bibfnamefont {Y.}~\bibnamefont
			{Guo}}, \bibinfo {author} {\bibfnamefont {J.}~\bibnamefont {Pack}}, \bibinfo
		{author} {\bibfnamefont {J.}~\bibnamefont {Swann}}, \bibinfo {author}
		{\bibfnamefont {L.}~\bibnamefont {Holtzman}}, \bibinfo {author}
		{\bibfnamefont {M.}~\bibnamefont {Cothrine}}, \bibinfo {author}
		{\bibfnamefont {K.}~\bibnamefont {Watanabe}}, \bibinfo {author}
		{\bibfnamefont {T.}~\bibnamefont {Taniguchi}}, \bibinfo {author}
		{\bibfnamefont {D.~G.}\ \bibnamefont {Mandrus}}, \bibinfo {author}
		{\bibfnamefont {K.}~\bibnamefont {Barmak}}, \bibinfo {author} {\bibfnamefont
			{J.}~\bibnamefont {Hone}}, \bibinfo {author} {\bibfnamefont {A.~J.}\
			\bibnamefont {Millis}}, \bibinfo {author} {\bibfnamefont {A.}~\bibnamefont
			{Pasupathy}}, \ and\ \bibinfo {author} {\bibfnamefont {C.~R.}\ \bibnamefont
			{Dean}},\ }\bibfield  {title} {\emph {\bibinfo {title} {{Superconductivity in
					$5.0^\circ$ twisted bilayer WSe$_2$}},\ }}\href {\doibase
		10.1038/s41586-024-08381-1} {\bibfield  {journal} {\bibinfo  {journal}
			{Nature}\ }\textbf {\bibinfo {volume} {637}},\ \bibinfo {pages} {839}
		(\bibinfo {year} {2025})}\BibitemShut {NoStop}%
	\bibitem [{\citenamefont {Xia}\ \emph {et~al.}(2025)\citenamefont {Xia},
		\citenamefont {Han}, \citenamefont {Watanabe}, \citenamefont {Taniguchi},
		\citenamefont {Shan},\ and\ \citenamefont {Mak}}]{KFMak_WSe2_2024}%
	\BibitemOpen
	\bibfield  {author} {\bibinfo {author} {\bibfnamefont {Y.}~\bibnamefont
			{Xia}}, \bibinfo {author} {\bibfnamefont {Z.}~\bibnamefont {Han}}, \bibinfo
		{author} {\bibfnamefont {K.}~\bibnamefont {Watanabe}}, \bibinfo {author}
		{\bibfnamefont {T.}~\bibnamefont {Taniguchi}}, \bibinfo {author}
		{\bibfnamefont {J.}~\bibnamefont {Shan}}, \ and\ \bibinfo {author}
		{\bibfnamefont {K.~F.}\ \bibnamefont {Mak}},\ }\bibfield  {title} {\emph
		{\bibinfo {title} {Superconductivity in twisted bilayer {{WSe$_2$}}},\
	}}\href {\doibase 10.1038/s41586-024-08116-2} {\bibfield  {journal} {\bibinfo
			{journal} {Nature}\ }\textbf {\bibinfo {volume} {637}},\ \bibinfo {pages}
		{833} (\bibinfo {year} {2025})}\BibitemShut {NoStop}%
	\bibitem [{\citenamefont {Fischer}\ \emph {et~al.}()\citenamefont {Fischer},
		\citenamefont {Klebl}, \citenamefont {Crépel}, \citenamefont {Ryee},
		\citenamefont {Rubio}, \citenamefont {Xian}, \citenamefont {Wehling},
		\citenamefont {Georges}, \citenamefont {Kennes},\ and\ \citenamefont
		{Millis}}]{Fischer_tWSe2_SC_2024}%
	\BibitemOpen
	\bibfield  {author} {\bibinfo {author} {\bibfnamefont {A.}~\bibnamefont
			{Fischer}}, \bibinfo {author} {\bibfnamefont {L.}~\bibnamefont {Klebl}},
		\bibinfo {author} {\bibfnamefont {V.}~\bibnamefont {Crépel}}, \bibinfo
		{author} {\bibfnamefont {S.}~\bibnamefont {Ryee}}, \bibinfo {author}
		{\bibfnamefont {A.}~\bibnamefont {Rubio}}, \bibinfo {author} {\bibfnamefont
			{L.}~\bibnamefont {Xian}}, \bibinfo {author} {\bibfnamefont {T.~O.}\
			\bibnamefont {Wehling}}, \bibinfo {author} {\bibfnamefont {A.}~\bibnamefont
			{Georges}}, \bibinfo {author} {\bibfnamefont {D.~M.}\ \bibnamefont {Kennes}},
		\ and\ \bibinfo {author} {\bibfnamefont {A.~J.}\ \bibnamefont {Millis}},\
	}\bibfield  {title} {\emph {\bibinfo {title} {{Theory of intervalley-coherent
					AFM order and topological superconductivity in tWSe$_2$}},\ }}\href
	{https://doi.org/10.48550/arXiv.2412.14296} {\ }\Eprint
	{http://arxiv.org/abs/arXiv:2412.14296}{arXiv:2412.14296}\BibitemShut
	{NoStop}%
	\bibitem [{\citenamefont {Schrade}\ and\ \citenamefont
		{Fu}(2024)}]{Schrade_tWSe2_SC_2024}%
	\BibitemOpen
	\bibfield  {author} {\bibinfo {author} {\bibfnamefont {C.}~\bibnamefont
			{Schrade}}\ and\ \bibinfo {author} {\bibfnamefont {L.}~\bibnamefont {Fu}},\
	}\bibfield  {title} {\emph {\bibinfo {title} {Nematic, chiral, and
				topological superconductivity in twisted transition metal dichalcogenides},\
	}}\href {\doibase 10.1103/PhysRevB.110.035143} {\bibfield  {journal}
		{\bibinfo  {journal} {Phys. Rev. B}\ }\textbf {\bibinfo {volume} {110}},\
		\bibinfo {pages} {035143} (\bibinfo {year} {2024})}\BibitemShut {NoStop}%
	\bibitem [{\citenamefont {Klebl}\ \emph {et~al.}(2023)\citenamefont {Klebl},
		\citenamefont {Fischer}, \citenamefont {Classen}, \citenamefont {Scherer},\
		and\ \citenamefont {Kennes}}]{Klebl_tWSe2_SC_2023}%
	\BibitemOpen
	\bibfield  {author} {\bibinfo {author} {\bibfnamefont {L.}~\bibnamefont
			{Klebl}}, \bibinfo {author} {\bibfnamefont {A.}~\bibnamefont {Fischer}},
		\bibinfo {author} {\bibfnamefont {L.}~\bibnamefont {Classen}}, \bibinfo
		{author} {\bibfnamefont {M.~M.}\ \bibnamefont {Scherer}}, \ and\ \bibinfo
		{author} {\bibfnamefont {D.~M.}\ \bibnamefont {Kennes}},\ }\bibfield  {title}
	{\emph {\bibinfo {title} {{Competition of density waves and superconductivity
					in twisted tungsten diselenide}},\ }}\href {\doibase
		10.1103/PhysRevResearch.5.L012034} {\bibfield  {journal} {\bibinfo  {journal}
			{Phys. Rev. Res.}\ }\textbf {\bibinfo {volume} {5}},\ \bibinfo {pages}
		{L012034} (\bibinfo {year} {2023})}\BibitemShut {NoStop}%
	\bibitem [{\citenamefont {Chubukov}\ and\ \citenamefont
		{Varma}(2025)}]{Chubukov_tWSe2_SC_2024}%
	\BibitemOpen
	\bibfield  {author} {\bibinfo {author} {\bibfnamefont {A.~V.}\ \bibnamefont
			{Chubukov}}\ and\ \bibinfo {author} {\bibfnamefont {C.~M.}\ \bibnamefont
			{Varma}},\ }\bibfield  {title} {\emph {\bibinfo {title} {Quantum criticality
				and superconductivity in twisted transition metal dichalcogenides},\ }}\href
	{\doibase 10.1103/PhysRevB.111.014507} {\bibfield  {journal} {\bibinfo
			{journal} {Phys. Rev. B}\ }\textbf {\bibinfo {volume} {111}},\ \bibinfo
		{pages} {014507} (\bibinfo {year} {2025})}\BibitemShut {NoStop}%
	\bibitem [{\citenamefont {Guerci}\ \emph {et~al.}()\citenamefont {Guerci},
		\citenamefont {Kaplan}, \citenamefont {Ingham}, \citenamefont {Pixley},\ and\
		\citenamefont {Millis}}]{Guerci_tWSe2_SC_2024}%
	\BibitemOpen
	\bibfield  {author} {\bibinfo {author} {\bibfnamefont {D.}~\bibnamefont
			{Guerci}}, \bibinfo {author} {\bibfnamefont {D.}~\bibnamefont {Kaplan}},
		\bibinfo {author} {\bibfnamefont {J.}~\bibnamefont {Ingham}}, \bibinfo
		{author} {\bibfnamefont {J.~H.}\ \bibnamefont {Pixley}}, \ and\ \bibinfo
		{author} {\bibfnamefont {A.~J.}\ \bibnamefont {Millis}},\ }\bibfield  {title}
	{\emph {\bibinfo {title} {{Topological superconductivity from repulsive
					interactions in twisted WSe$_2$}},\ }}\href
	{https://doi.org/10.48550/arXiv.2408.16075} {\ }\Eprint
	{http://arxiv.org/abs/arXiv:2408.16075}{arXiv:2408.16075}\BibitemShut
	{NoStop}%
	\bibitem [{\citenamefont {Qin}\ \emph {et~al.}()\citenamefont {Qin},
		\citenamefont {Qiu},\ and\ \citenamefont {Wu}}]{Qin_tWSe2_SC_2024}%
	\BibitemOpen
	\bibfield  {author} {\bibinfo {author} {\bibfnamefont {W.}~\bibnamefont
			{Qin}}, \bibinfo {author} {\bibfnamefont {W.-X.}\ \bibnamefont {Qiu}}, \ and\
		\bibinfo {author} {\bibfnamefont {F.}~\bibnamefont {Wu}},\ }\bibfield
	{title} {\emph {\bibinfo {title} {{Kohn-Luttinger Mechanism of
					Superconductivity in Twisted Bilayer WSe$_2$: Gate-Tunable Unconventional
					Pairing Symmetry}},\ }}\href {https://doi.org/10.48550/arXiv.2409.16114} {\
	}\Eprint
	{http://arxiv.org/abs/arXiv:2409.16114}{arXiv:2409.16114}\BibitemShut
	{NoStop}%
	\bibitem [{\citenamefont {Zhu}\ \emph {et~al.}(2025)\citenamefont {Zhu},
		\citenamefont {Chou}, \citenamefont {Xie},\ and\ \citenamefont
		{Das~Sarma}}]{JZhu_WSe2_SC_2025}%
	\BibitemOpen
	\bibfield  {author} {\bibinfo {author} {\bibfnamefont {J.}~\bibnamefont
			{Zhu}}, \bibinfo {author} {\bibfnamefont {Y.-Z.}\ \bibnamefont {Chou}},
		\bibinfo {author} {\bibfnamefont {M.}~\bibnamefont {Xie}}, \ and\ \bibinfo
		{author} {\bibfnamefont {S.}~\bibnamefont {Das~Sarma}},\ }\bibfield  {title}
	{\emph {\bibinfo {title} {Superconductivity in twisted transition metal
				dichalcogenide homobilayers},\ }}\href {\doibase
		10.1103/PhysRevB.111.L060501} {\bibfield  {journal} {\bibinfo  {journal}
			{Phys. Rev. B}\ }\textbf {\bibinfo {volume} {111}},\ \bibinfo {pages}
		{L060501} (\bibinfo {year} {2025})}\BibitemShut {NoStop}%
	\bibitem [{\citenamefont {Chuyi}\ \emph {et~al.}()\citenamefont {Chuyi},
		\citenamefont {Ming-Rui}, \citenamefont {Wu}, \citenamefont {Sun},\ and\
		\citenamefont {Yao}}]{Tuo_tWSe2_SC_2024}%
	\BibitemOpen
	\bibfield  {author} {\bibinfo {author} {\bibfnamefont {T.}~\bibnamefont
			{Chuyi}}, \bibinfo {author} {\bibfnamefont {L.}~\bibnamefont {Ming-Rui}},
		\bibinfo {author} {\bibfnamefont {Z.}~\bibnamefont {Wu}}, \bibinfo {author}
		{\bibfnamefont {W.}~\bibnamefont {Sun}}, \ and\ \bibinfo {author}
		{\bibfnamefont {H.}~\bibnamefont {Yao}},\ }\bibfield  {title} {\emph
		{\bibinfo {title} {{Theory of Topological Superconductivity and
					Antiferromagnetic Correlated Insulators in Twisted Bilayer WSe$_2$}},\
	}}\href {https://doi.org/10.48550/arXiv.2409.06779} {\ }\Eprint
	{http://arxiv.org/abs/arXiv:2409.06779}{arXiv:2409.06779}\BibitemShut
	{NoStop}%
	\bibitem [{\citenamefont {Kim}\ \emph {et~al.}(2025{\natexlab{a}})\citenamefont
		{Kim}, \citenamefont {Mendez-Valderrama}, \citenamefont {Wang},\ and\
		\citenamefont {Chowdhury}}]{Kim_tWSe2_SC_2025}%
	\BibitemOpen
	\bibfield  {author} {\bibinfo {author} {\bibfnamefont {S.}~\bibnamefont
			{Kim}}, \bibinfo {author} {\bibfnamefont {J.~F.}\ \bibnamefont
			{Mendez-Valderrama}}, \bibinfo {author} {\bibfnamefont {X.}~\bibnamefont
			{Wang}}, \ and\ \bibinfo {author} {\bibfnamefont {D.}~\bibnamefont
			{Chowdhury}},\ }\bibfield  {title} {\emph {\bibinfo {title} {{Theory of
					correlated insulators and superconductor at $\nu = 1$ in twisted WSe$_2$}},\
	}}\href {\doibase 10.1038/s41467-025-56816-8} {\bibfield  {journal} {\bibinfo
			{journal} {Nature Communications}\ }\textbf {\bibinfo {volume} {16}},\
		\bibinfo {pages} {1701} (\bibinfo {year} {2025}{\natexlab{a}})}\BibitemShut
	{NoStop}%
	\bibitem [{\citenamefont {Wietek}\ \emph {et~al.}(2022)\citenamefont {Wietek},
		\citenamefont {Wang}, \citenamefont {Zang}, \citenamefont {Cano},
		\citenamefont {Georges},\ and\ \citenamefont {Millis}}]{Wietek_tTMD_SC_2022}%
	\BibitemOpen
	\bibfield  {author} {\bibinfo {author} {\bibfnamefont {A.}~\bibnamefont
			{Wietek}}, \bibinfo {author} {\bibfnamefont {J.}~\bibnamefont {Wang}},
		\bibinfo {author} {\bibfnamefont {J.}~\bibnamefont {Zang}}, \bibinfo {author}
		{\bibfnamefont {J.}~\bibnamefont {Cano}}, \bibinfo {author} {\bibfnamefont
			{A.}~\bibnamefont {Georges}}, \ and\ \bibinfo {author} {\bibfnamefont
			{A.}~\bibnamefont {Millis}},\ }\bibfield  {title} {\emph {\bibinfo {title}
			{{Tunable stripe order and weak superconductivity in the Moir\'e Hubbard
					model}},\ }}\href {\doibase 10.1103/PhysRevResearch.4.043048} {\bibfield
		{journal} {\bibinfo  {journal} {Phys. Rev. Res.}\ }\textbf {\bibinfo {volume}
			{4}},\ \bibinfo {pages} {043048} (\bibinfo {year} {2022})}\BibitemShut
	{NoStop}%
	\bibitem [{\citenamefont {Christos}\ \emph
		{et~al.}({\natexlab{a}})\citenamefont {Christos}, \citenamefont {Bonetti},\
		and\ \citenamefont {Scheurer}}]{Christos_tWSe2_SC_2024}%
	\BibitemOpen
	\bibfield  {author} {\bibinfo {author} {\bibfnamefont {M.}~\bibnamefont
			{Christos}}, \bibinfo {author} {\bibfnamefont {P.~M.}\ \bibnamefont
			{Bonetti}}, \ and\ \bibinfo {author} {\bibfnamefont {M.~S.}\ \bibnamefont
			{Scheurer}},\ }\bibfield  {title} {\emph {\bibinfo {title} {{Approximate
					symmetries, insulators, and superconductivity in continuum-model description
					of twisted WSe$_2$}},\ }}\href {https://doi.org/10.48550/arXiv.2407.02393} {\
		({\natexlab{a}})},\ \Eprint
	{http://arxiv.org/abs/arXiv:2407.02393}{arXiv:2407.02393}\BibitemShut
	{NoStop}%
	\bibitem [{\citenamefont {Xie}\ \emph {et~al.}(2025)\citenamefont {Xie},
		\citenamefont {Chen}, \citenamefont {Sur}, \citenamefont {Fang},
		\citenamefont {Cano},\ and\ \citenamefont {Si}}]{Xie_tWSe2_SC_2024}%
	\BibitemOpen
	\bibfield  {author} {\bibinfo {author} {\bibfnamefont {F.}~\bibnamefont
			{Xie}}, \bibinfo {author} {\bibfnamefont {L.}~\bibnamefont {Chen}}, \bibinfo
		{author} {\bibfnamefont {S.}~\bibnamefont {Sur}}, \bibinfo {author}
		{\bibfnamefont {Y.}~\bibnamefont {Fang}}, \bibinfo {author} {\bibfnamefont
			{J.}~\bibnamefont {Cano}}, \ and\ \bibinfo {author} {\bibfnamefont
			{Q.}~\bibnamefont {Si}},\ }\bibfield  {title} {\emph {\bibinfo {title}
			{Superconductivity in twisted ${\mathrm{wse}}_{2}$ from topology-induced
				quantum fluctuations},\ }}\href {\doibase 10.1103/PhysRevLett.134.136503}
	{\bibfield  {journal} {\bibinfo  {journal} {Phys. Rev. Lett.}\ }\textbf
		{\bibinfo {volume} {134}},\ \bibinfo {pages} {136503} (\bibinfo {year}
		{2025})}\BibitemShut {NoStop}%
	\bibitem [{\citenamefont {Le}\ \emph {et~al.}(2015)\citenamefont {Le},
		\citenamefont {Barinov}, \citenamefont {Preciado}, \citenamefont {Isarraraz},
		\citenamefont {Tanabe}, \citenamefont {Komesu}, \citenamefont {Troha},
		\citenamefont {Bartels}, \citenamefont {Rahman},\ and\ \citenamefont
		{Dowben}}]{Le_WSe2_SOC_2015}%
	\BibitemOpen
	\bibfield  {author} {\bibinfo {author} {\bibfnamefont {D.}~\bibnamefont
			{Le}}, \bibinfo {author} {\bibfnamefont {A.}~\bibnamefont {Barinov}},
		\bibinfo {author} {\bibfnamefont {E.}~\bibnamefont {Preciado}}, \bibinfo
		{author} {\bibfnamefont {M.}~\bibnamefont {Isarraraz}}, \bibinfo {author}
		{\bibfnamefont {I.}~\bibnamefont {Tanabe}}, \bibinfo {author} {\bibfnamefont
			{T.}~\bibnamefont {Komesu}}, \bibinfo {author} {\bibfnamefont
			{C.}~\bibnamefont {Troha}}, \bibinfo {author} {\bibfnamefont
			{L.}~\bibnamefont {Bartels}}, \bibinfo {author} {\bibfnamefont {T.~S.}\
			\bibnamefont {Rahman}}, \ and\ \bibinfo {author} {\bibfnamefont {P.~A.}\
			\bibnamefont {Dowben}},\ }\bibfield  {title} {\emph {\bibinfo {title}
			{{Spin–orbit coupling in the band structure of monolayer WSe$_2$}},\
	}}\href {\doibase 10.1088/0953-8984/27/18/182201} {\bibfield  {journal}
		{\bibinfo  {journal} {Journal of Physics: Condensed Matter}\ }\textbf
		{\bibinfo {volume} {27}},\ \bibinfo {pages} {182201} (\bibinfo {year}
		{2015})}\BibitemShut {NoStop}%
	\bibitem [{\citenamefont {Fulde}\ and\ \citenamefont
		{Ferrell}(1964)}]{FF_1964}%
	\BibitemOpen
	\bibfield  {author} {\bibinfo {author} {\bibfnamefont {P.}~\bibnamefont
			{Fulde}}\ and\ \bibinfo {author} {\bibfnamefont {R.~A.}\ \bibnamefont
			{Ferrell}},\ }\bibfield  {title} {\emph {\bibinfo {title} {Superconductivity
				in a strong spin-exchange field},\ }}\href {\doibase
		10.1103/PhysRev.135.A550} {\bibfield  {journal} {\bibinfo  {journal} {Phys.
				Rev.}\ }\textbf {\bibinfo {volume} {135}},\ \bibinfo {pages} {A550} (\bibinfo
		{year} {1964})}\BibitemShut {NoStop}%
	\bibitem [{\citenamefont {Larkin}\ and\ \citenamefont
		{Ovchinnikov}(1964)}]{LO_1965}%
	\BibitemOpen
	\bibfield  {author} {\bibinfo {author} {\bibfnamefont {A.~I.}\ \bibnamefont
			{Larkin}}\ and\ \bibinfo {author} {\bibfnamefont {Y.~N.}\ \bibnamefont
			{Ovchinnikov}},\ }\bibfield  {title} {\emph {\bibinfo {title} {Nonuniform
				state of superconductors},\ }}\href@noop {} {\bibfield  {journal} {\bibinfo
			{journal} {Zh. Eksp. Teor. Fiz.}\ }\textbf {\bibinfo {volume} {47}},\
		\bibinfo {pages} {1136} (\bibinfo {year} {1964})},\ \bibinfo {note} {english
		translation: Sov. Phys. JETP \textbf{20}, 762 (1965)}\BibitemShut {NoStop}%
	\bibitem [{\citenamefont {Wan}\ \emph {et~al.}(2023)\citenamefont {Wan},
		\citenamefont {Zheliuk}, \citenamefont {Yuan}, \citenamefont {Peng},
		\citenamefont {Zhang}, \citenamefont {Liang}, \citenamefont {Zeitler},
		\citenamefont {Wiedmann}, \citenamefont {Hussey}, \citenamefont {Palstra},\
		and\ \citenamefont {Ye}}]{PWan_obitalFFLO_2023}%
	\BibitemOpen
	\bibfield  {author} {\bibinfo {author} {\bibfnamefont {P.}~\bibnamefont
			{Wan}}, \bibinfo {author} {\bibfnamefont {O.}~\bibnamefont {Zheliuk}},
		\bibinfo {author} {\bibfnamefont {N.~F.~Q.}\ \bibnamefont {Yuan}}, \bibinfo
		{author} {\bibfnamefont {X.}~\bibnamefont {Peng}}, \bibinfo {author}
		{\bibfnamefont {L.}~\bibnamefont {Zhang}}, \bibinfo {author} {\bibfnamefont
			{M.}~\bibnamefont {Liang}}, \bibinfo {author} {\bibfnamefont
			{U.}~\bibnamefont {Zeitler}}, \bibinfo {author} {\bibfnamefont
			{S.}~\bibnamefont {Wiedmann}}, \bibinfo {author} {\bibfnamefont {N.~E.}\
			\bibnamefont {Hussey}}, \bibinfo {author} {\bibfnamefont {T.~T.~M.}\
			\bibnamefont {Palstra}}, \ and\ \bibinfo {author} {\bibfnamefont
			{J.}~\bibnamefont {Ye}},\ }\bibfield  {title} {\emph {\bibinfo {title}
			{{Orbital Fulde–Ferrell–Larkin–Ovchinnikov state in an Ising
					superconductor}},\ }}\href {\doibase 10.1038/s41586-023-05967-z} {\bibfield
		{journal} {\bibinfo  {journal} {Nature}\ }\textbf {\bibinfo {volume} {619}},\
		\bibinfo {pages} {46} (\bibinfo {year} {2023})}\BibitemShut {NoStop}%
	\bibitem [{\citenamefont {Liu}(2017)}]{CXLiu_FFLO_orb_2017}%
	\BibitemOpen
	\bibfield  {author} {\bibinfo {author} {\bibfnamefont {C.-X.}\ \bibnamefont
			{Liu}},\ }\bibfield  {title} {\emph {\bibinfo {title} {Unconventional
				superconductivity in bilayer transition metal dichalcogenides},\ }}\href
	{\doibase 10.1103/PhysRevLett.118.087001} {\bibfield  {journal} {\bibinfo
			{journal} {Phys. Rev. Lett.}\ }\textbf {\bibinfo {volume} {118}},\ \bibinfo
		{pages} {087001} (\bibinfo {year} {2017})}\BibitemShut {NoStop}%
	\bibitem [{\citenamefont {Xie}\ and\ \citenamefont
		{Law}(2023)}]{YMXie_FFLO_2023}%
	\BibitemOpen
	\bibfield  {author} {\bibinfo {author} {\bibfnamefont {Y.-M.}\ \bibnamefont
			{Xie}}\ and\ \bibinfo {author} {\bibfnamefont {K.~T.}\ \bibnamefont {Law}},\
	}\bibfield  {title} {\emph {\bibinfo {title} {Orbital {F}ulde-{F}errell
				pairing state in moir\'e {I}sing superconductors},\ }}\href {\doibase
		10.1103/PhysRevLett.131.016001} {\bibfield  {journal} {\bibinfo  {journal}
			{Phys. Rev. Lett.}\ }\textbf {\bibinfo {volume} {131}},\ \bibinfo {pages}
		{016001} (\bibinfo {year} {2023})}\BibitemShut {NoStop}%
	\bibitem [{\citenamefont {Yuan}(2023)}]{Yuan_orbitalFFLO_2023}%
	\BibitemOpen
	\bibfield  {author} {\bibinfo {author} {\bibfnamefont {N.~F.~Q.}\
			\bibnamefont {Yuan}},\ }\bibfield  {title} {\emph {\bibinfo {title} {{Orbital
					Fulde-Ferrell-Larkin-Ovchinnikov state in an Ising superconductor}},\ }}\href
	{\doibase 10.1103/PhysRevResearch.5.043122} {\bibfield  {journal} {\bibinfo
			{journal} {Phys. Rev. Res.}\ }\textbf {\bibinfo {volume} {5}},\ \bibinfo
		{pages} {043122} (\bibinfo {year} {2023})}\BibitemShut {NoStop}%
	\bibitem [{\citenamefont {Yuan}()}]{Yuan_orbitalFFLO_2025}%
	\BibitemOpen
	\bibfield  {author} {\bibinfo {author} {\bibfnamefont {N.~F.~Q.}\
			\bibnamefont {Yuan}},\ }\bibfield  {title} {\emph {\bibinfo {title} {{Orbital
					Fulde-Ferrell State versus Orbital Larkin-Ovchinnikov State}},\ }}\href
	{https://doi.org/10.48550/arXiv.2502.18075} {\ }\Eprint
	{http://arxiv.org/abs/arXiv:2502.18075}{arXiv:2502.18075}\BibitemShut
	{NoStop}%
	\bibitem [{\citenamefont {Zhao}\ \emph {et~al.}(2023)\citenamefont {Zhao},
		\citenamefont {Debbeler}, \citenamefont {Kühne}, \citenamefont {Fecher},
		\citenamefont {Gross},\ and\ \citenamefont {Smet}}]{Zhao_orbitalFFLO_2023}%
	\BibitemOpen
	\bibfield  {author} {\bibinfo {author} {\bibfnamefont {D.}~\bibnamefont
			{Zhao}}, \bibinfo {author} {\bibfnamefont {L.}~\bibnamefont {Debbeler}},
		\bibinfo {author} {\bibfnamefont {M.}~\bibnamefont {Kühne}}, \bibinfo
		{author} {\bibfnamefont {S.}~\bibnamefont {Fecher}}, \bibinfo {author}
		{\bibfnamefont {N.}~\bibnamefont {Gross}}, \ and\ \bibinfo {author}
		{\bibfnamefont {J.}~\bibnamefont {Smet}},\ }\bibfield  {title} {\emph
		{\bibinfo {title} {Evidence of finite-momentum pairing in a centrosymmetric
				bilayer},\ }}\href {\doibase 10.1038/s41567-023-02202-4} {\bibfield
		{journal} {\bibinfo  {journal} {Nature Physics}\ }\textbf {\bibinfo {volume}
			{19}},\ \bibinfo {pages} {1599} (\bibinfo {year} {2023})}\BibitemShut
	{NoStop}%
	\bibitem [{\citenamefont {Cao}\ \emph {et~al.}()\citenamefont {Cao},
		\citenamefont {Liao}, \citenamefont {Yan}, \citenamefont {Zhu}, \citenamefont
		{Liguo}, \citenamefont {Watanabe}, \citenamefont {Taniguchi}, \citenamefont
		{Morpurgo}, \citenamefont {Liu}, \citenamefont {Xue},\ and\ \citenamefont
		{Zhang}}]{Cao_orbitalFFLO_2024}%
	\BibitemOpen
	\bibfield  {author} {\bibinfo {author} {\bibfnamefont {Z.}~\bibnamefont
			{Cao}}, \bibinfo {author} {\bibfnamefont {M.}~\bibnamefont {Liao}}, \bibinfo
		{author} {\bibfnamefont {H.}~\bibnamefont {Yan}}, \bibinfo {author}
		{\bibfnamefont {Y.}~\bibnamefont {Zhu}}, \bibinfo {author} {\bibfnamefont
			{Z.}~\bibnamefont {Liguo}}, \bibinfo {author} {\bibfnamefont
			{K.}~\bibnamefont {Watanabe}}, \bibinfo {author} {\bibfnamefont
			{T.}~\bibnamefont {Taniguchi}}, \bibinfo {author} {\bibfnamefont {A.~F.}\
			\bibnamefont {Morpurgo}}, \bibinfo {author} {\bibfnamefont {H.}~\bibnamefont
			{Liu}}, \bibinfo {author} {\bibfnamefont {Q.-K.}\ \bibnamefont {Xue}}, \ and\
		\bibinfo {author} {\bibfnamefont {D.}~\bibnamefont {Zhang}},\ }\bibfield
	{title} {\emph {\bibinfo {title} {{Spectroscopic evidence for a first-order
					transition to the orbital Fulde-Ferrell-Larkin-Ovchinnikov state}},\ }}\href
	{https://doi.org/10.48550/arXiv.2409.00373} {\ }\Eprint
	{http://arxiv.org/abs/arXiv:2409.00373}{arXiv:2409.00373}\BibitemShut
	{NoStop}%
	\bibitem [{\citenamefont {Zhao}\ \emph {et~al.}()\citenamefont {Zhao},
		\citenamefont {Guo}, \citenamefont {Yan}, \citenamefont {Yuan}, \citenamefont
		{Zhao}, \citenamefont {Guan}, \citenamefont {Lan}, \citenamefont {Li},
		\citenamefont {Liu},\ and\ \citenamefont {Wang}}]{Zhao_orbitalFFLO_2024}%
	\BibitemOpen
	\bibfield  {author} {\bibinfo {author} {\bibfnamefont {X.}~\bibnamefont
			{Zhao}}, \bibinfo {author} {\bibfnamefont {G.}~\bibnamefont {Guo}}, \bibinfo
		{author} {\bibfnamefont {C.}~\bibnamefont {Yan}}, \bibinfo {author}
		{\bibfnamefont {N.~F.}\ \bibnamefont {Yuan}}, \bibinfo {author}
		{\bibfnamefont {C.}~\bibnamefont {Zhao}}, \bibinfo {author} {\bibfnamefont
			{H.}~\bibnamefont {Guan}}, \bibinfo {author} {\bibfnamefont {C.}~\bibnamefont
			{Lan}}, \bibinfo {author} {\bibfnamefont {Y.}~\bibnamefont {Li}}, \bibinfo
		{author} {\bibfnamefont {X.}~\bibnamefont {Liu}}, \ and\ \bibinfo {author}
		{\bibfnamefont {S.}~\bibnamefont {Wang}},\ }\bibfield  {title} {\emph
		{\bibinfo {title} {{Orbital Fulde-Ferrell-Larkin-Ovchinnikov state in
					$2H$-$NbS_2$ flakes}},\ }}\href {https://doi.org/10.48550/arXiv.2411.08980}
	{\ }\Eprint
	{http://arxiv.org/abs/arXiv:2411.08980}{arXiv:2411.08980}\BibitemShut
	{NoStop}%
	\bibitem [{\citenamefont {Cho}\ \emph {et~al.}()\citenamefont {Cho},
		\citenamefont {Lortz}, \citenamefont {Lo}, \citenamefont {Ng}, \citenamefont
		{Chui}, \citenamefont {Allan}, \citenamefont {Abdel-Hafiez}, \citenamefont
		{Park}, \citenamefont {Cho}, \citenamefont {Park}, \citenamefont {Yuan},\
		and\ \citenamefont {Lortz}}]{Cho_orbitalFFLO_2024}%
	\BibitemOpen
	\bibfield  {author} {\bibinfo {author} {\bibfnamefont {C.-w.}\ \bibnamefont
			{Cho}}, \bibinfo {author} {\bibfnamefont {T.~T.}\ \bibnamefont {Lortz}},
		\bibinfo {author} {\bibfnamefont {K.~T.}\ \bibnamefont {Lo}}, \bibinfo
		{author} {\bibfnamefont {C.~Y.}\ \bibnamefont {Ng}}, \bibinfo {author}
		{\bibfnamefont {S.~H.}\ \bibnamefont {Chui}}, \bibinfo {author}
		{\bibfnamefont {A.~R.}\ \bibnamefont {Allan}}, \bibinfo {author}
		{\bibfnamefont {M.}~\bibnamefont {Abdel-Hafiez}}, \bibinfo {author}
		{\bibfnamefont {J.}~\bibnamefont {Park}}, \bibinfo {author} {\bibfnamefont
			{B.}~\bibnamefont {Cho}}, \bibinfo {author} {\bibfnamefont {K.}~\bibnamefont
			{Park}}, \bibinfo {author} {\bibfnamefont {N.~F.~Q.}\ \bibnamefont {Yuan}}, \
		and\ \bibinfo {author} {\bibfnamefont {R.}~\bibnamefont {Lortz}},\ }\bibfield
	{title} {\emph {\bibinfo {title} {{Evidence for the novel type of orbital
					Fulde-Ferrell-Larkin-Ovchinnikov state in the bulk limit of $2H$-$NbSe_2$}},\
	}}\href {https://doi.org/10.48550/arXiv.2312.03215} {\ }\Eprint
	{http://arxiv.org/abs/arXiv:2312.03215}{arXiv:2312.03215}\BibitemShut
	{NoStop}%
	\bibitem [{\citenamefont {Yang}\ \emph {et~al.}()\citenamefont {Yang},
		\citenamefont {Zhang}, \citenamefont {Mandal}, \citenamefont {Meng},
		\citenamefont {Fabbris}, \citenamefont {Said}, \citenamefont {Lozano},
		\citenamefont {Rajapitamahuni}, \citenamefont {Vescovo}, \citenamefont
		{Nelson}, \citenamefont {Lin}, \citenamefont {Park}, \citenamefont
		{Clements}, \citenamefont {Ward}, \citenamefont {Lee}, \citenamefont {Lei},
		\citenamefont {Liu},\ and\ \citenamefont {Miao}}]{Yang_orbitalFFLO_2024}%
	\BibitemOpen
	\bibfield  {author} {\bibinfo {author} {\bibfnamefont {F.~Z.}\ \bibnamefont
			{Yang}}, \bibinfo {author} {\bibfnamefont {H.~D.}\ \bibnamefont {Zhang}},
		\bibinfo {author} {\bibfnamefont {S.}~\bibnamefont {Mandal}}, \bibinfo
		{author} {\bibfnamefont {F.~Y.}\ \bibnamefont {Meng}}, \bibinfo {author}
		{\bibfnamefont {G.}~\bibnamefont {Fabbris}}, \bibinfo {author} {\bibfnamefont
			{A.}~\bibnamefont {Said}}, \bibinfo {author} {\bibfnamefont {P.~M.}\
			\bibnamefont {Lozano}}, \bibinfo {author} {\bibfnamefont {A.}~\bibnamefont
			{Rajapitamahuni}}, \bibinfo {author} {\bibfnamefont {E.}~\bibnamefont
			{Vescovo}}, \bibinfo {author} {\bibfnamefont {C.}~\bibnamefont {Nelson}},
		\bibinfo {author} {\bibfnamefont {S.}~\bibnamefont {Lin}}, \bibinfo {author}
		{\bibfnamefont {Y.}~\bibnamefont {Park}}, \bibinfo {author} {\bibfnamefont
			{E.~M.}\ \bibnamefont {Clements}}, \bibinfo {author} {\bibfnamefont {T.~Z.}\
			\bibnamefont {Ward}}, \bibinfo {author} {\bibfnamefont {H.-N.}\ \bibnamefont
			{Lee}}, \bibinfo {author} {\bibfnamefont {H.~C.}\ \bibnamefont {Lei}},
		\bibinfo {author} {\bibfnamefont {C.~X.}\ \bibnamefont {Liu}}, \ and\
		\bibinfo {author} {\bibfnamefont {H.}~\bibnamefont {Miao}},\ }\bibfield
	{title} {\emph {\bibinfo {title} {{Signature of Orbital Driven Finite
					Momentum Pairing in a 3D Ising Superconductor}},\ }}\href
	{https://doi.org/10.48550/arXiv.2407.10352} {\ }\Eprint
	{http://arxiv.org/abs/arXiv:2407.10352}{arXiv:2407.10352}\BibitemShut
	{NoStop}%
	\bibitem [{\citenamefont {Qiu}\ and\ \citenamefont
		{Zhou}(2022)}]{Qiu_orbitalFFLO_2022}%
	\BibitemOpen
	\bibfield  {author} {\bibinfo {author} {\bibfnamefont {G.-W.}\ \bibnamefont
			{Qiu}}\ and\ \bibinfo {author} {\bibfnamefont {Y.}~\bibnamefont {Zhou}},\
	}\bibfield  {title} {\emph {\bibinfo {title} {Inhomogeneous superconducting
				states in two weakly linked superconducting ultrathin films},\ }}\href
	{\doibase 10.1103/PhysRevB.105.L100506} {\bibfield  {journal} {\bibinfo
			{journal} {Phys. Rev. B}\ }\textbf {\bibinfo {volume} {105}},\ \bibinfo
		{pages} {L100506} (\bibinfo {year} {2022})}\BibitemShut {NoStop}%
	\bibitem [{\citenamefont {Yan}\ \emph {et~al.}()\citenamefont {Yan},
		\citenamefont {Liu}, \citenamefont {Liu}, \citenamefont {Zhang},\ and\
		\citenamefont {Xie}}]{Yan_orbitalFFLO_2024}%
	\BibitemOpen
	\bibfield  {author} {\bibinfo {author} {\bibfnamefont {H.}~\bibnamefont
			{Yan}}, \bibinfo {author} {\bibfnamefont {H.}~\bibnamefont {Liu}}, \bibinfo
		{author} {\bibfnamefont {Y.}~\bibnamefont {Liu}}, \bibinfo {author}
		{\bibfnamefont {D.}~\bibnamefont {Zhang}}, \ and\ \bibinfo {author}
		{\bibfnamefont {X.~C.}\ \bibnamefont {Xie}},\ }\bibfield  {title} {\emph
		{\bibinfo {title} {{Orbital-FFLO State and Josephson Vortex Lattice Melting
					in Layered Ising Superconductors}},\ }}\href
	{https://doi.org/10.48550/arXiv.2409.20336} {\ }\Eprint
	{http://arxiv.org/abs/arXiv:2409.20336}{arXiv:2409.20336}\BibitemShut
	{NoStop}%
	\bibitem [{\citenamefont {Nag}\ \emph {et~al.}()\citenamefont {Nag},
		\citenamefont {Schirmer}, \citenamefont {Rossi}, \citenamefont {Liu},\ and\
		\citenamefont {Jain}}]{Nag_orbitalFFLO_2024}%
	\BibitemOpen
	\bibfield  {author} {\bibinfo {author} {\bibfnamefont {U.}~\bibnamefont
			{Nag}}, \bibinfo {author} {\bibfnamefont {J.}~\bibnamefont {Schirmer}},
		\bibinfo {author} {\bibfnamefont {E.}~\bibnamefont {Rossi}}, \bibinfo
		{author} {\bibfnamefont {C.-X.}\ \bibnamefont {Liu}}, \ and\ \bibinfo
		{author} {\bibfnamefont {J.~K.}\ \bibnamefont {Jain}},\ }\bibfield  {title}
	{\emph {\bibinfo {title} {{BCS Stripe Phase in Coupled Bilayer
					Superconductors}},\ }}\href {https://doi.org/10.48550/arXiv.2408.00689} {\
	}\Eprint
	{http://arxiv.org/abs/arXiv:2408.00689}{arXiv:2408.00689}\BibitemShut
	{NoStop}%
	\bibitem [{\citenamefont {Chandrasekhar}(1962)}]{Chandrasekhar_1962}%
	\BibitemOpen
	\bibfield  {author} {\bibinfo {author} {\bibfnamefont {B.~S.}\ \bibnamefont
			{Chandrasekhar}},\ }\bibfield  {title} {\emph {\bibinfo {title} {A note on
				the maximum critical field of high-field superconductors},\ }}\href
	{https://www.osti.gov/biblio/4734493} {\bibfield  {journal} {\bibinfo
			{journal} {Appl. Phys. Letters}\ }\textbf {\bibinfo {volume} {Vol: 1}}
		(\bibinfo {year} {1962})}\BibitemShut {NoStop}%
	\bibitem [{\citenamefont {Clogston}(1962)}]{Clogston_1962}%
	\BibitemOpen
	\bibfield  {author} {\bibinfo {author} {\bibfnamefont {A.~M.}\ \bibnamefont
			{Clogston}},\ }\bibfield  {title} {\emph {\bibinfo {title} {Upper limit for
				the critical field in hard superconductors},\ }}\href {\doibase
		10.1103/PhysRevLett.9.266} {\bibfield  {journal} {\bibinfo  {journal} {Phys.
				Rev. Lett.}\ }\textbf {\bibinfo {volume} {9}},\ \bibinfo {pages} {266}
		(\bibinfo {year} {1962})}\BibitemShut {NoStop}%
	\bibitem [{\citenamefont {Agterberg}\ \emph {et~al.}(2020)\citenamefont
		{Agterberg}, \citenamefont {Davis}, \citenamefont {Edkins}, \citenamefont
		{Fradkin}, \citenamefont {Van~Harlingen}, \citenamefont {Kivelson},
		\citenamefont {Lee}, \citenamefont {Radzihovsky}, \citenamefont {Tranquada},\
		and\ \citenamefont {Wang}}]{Agterberg_PDW_2020}%
	\BibitemOpen
	\bibfield  {author} {\bibinfo {author} {\bibfnamefont {D.~F.}\ \bibnamefont
			{Agterberg}}, \bibinfo {author} {\bibfnamefont {J.~S.}\ \bibnamefont
			{Davis}}, \bibinfo {author} {\bibfnamefont {S.~D.}\ \bibnamefont {Edkins}},
		\bibinfo {author} {\bibfnamefont {E.}~\bibnamefont {Fradkin}}, \bibinfo
		{author} {\bibfnamefont {D.~J.}\ \bibnamefont {Van~Harlingen}}, \bibinfo
		{author} {\bibfnamefont {S.~A.}\ \bibnamefont {Kivelson}}, \bibinfo {author}
		{\bibfnamefont {P.~A.}\ \bibnamefont {Lee}}, \bibinfo {author} {\bibfnamefont
			{L.}~\bibnamefont {Radzihovsky}}, \bibinfo {author} {\bibfnamefont {J.~M.}\
			\bibnamefont {Tranquada}}, \ and\ \bibinfo {author} {\bibfnamefont
			{Y.}~\bibnamefont {Wang}},\ }\bibfield  {title} {\emph {\bibinfo {title} {The
				physics of pair-density waves: Cuprate superconductors and beyond},\ }}\href
	{\doibase https://doi.org/10.1146/annurev-conmatphys-031119-050711}
	{\bibfield  {journal} {\bibinfo  {journal} {Annual Review of Condensed Matter
				Physics}\ }\textbf {\bibinfo {volume} {11}},\ \bibinfo {pages} {231}
		(\bibinfo {year} {2020})}\BibitemShut {NoStop}%
	\bibitem [{\citenamefont {Casalbuoni}\ and\ \citenamefont
		{Nardulli}(2004)}]{Casalbuoni_FFLO_2004}%
	\BibitemOpen
	\bibfield  {author} {\bibinfo {author} {\bibfnamefont {R.}~\bibnamefont
			{Casalbuoni}}\ and\ \bibinfo {author} {\bibfnamefont {G.}~\bibnamefont
			{Nardulli}},\ }\bibfield  {title} {\emph {\bibinfo {title} {{Inhomogeneous
					superconductivity in condensed matter and QCD}},\ }}\href {\doibase
		10.1103/RevModPhys.76.263} {\bibfield  {journal} {\bibinfo  {journal} {Rev.
				Mod. Phys.}\ }\textbf {\bibinfo {volume} {76}},\ \bibinfo {pages} {263}
		(\bibinfo {year} {2004})}\BibitemShut {NoStop}%
	\bibitem [{\citenamefont {Radzihovsky}\ and\ \citenamefont
		{Sheehy}(2010)}]{Radzihovsky_2010}%
	\BibitemOpen
	\bibfield  {author} {\bibinfo {author} {\bibfnamefont {L.}~\bibnamefont
			{Radzihovsky}}\ and\ \bibinfo {author} {\bibfnamefont {D.~E.}\ \bibnamefont
			{Sheehy}},\ }\bibfield  {title} {\emph {\bibinfo {title} {{Imbalanced
					Feshbach-resonant Fermi gases}},\ }}\href {\doibase
		10.1088/0034-4885/73/7/076501} {\bibfield  {journal} {\bibinfo  {journal}
			{Reports on Progress in Physics}\ }\textbf {\bibinfo {volume} {73}},\
		\bibinfo {pages} {076501} (\bibinfo {year} {2010})}\BibitemShut {NoStop}%
	\bibitem [{\citenamefont {Houzet}\ \emph {et~al.}(2002)\citenamefont {Houzet},
		\citenamefont {Buzdin}, \citenamefont {Bulaevskii},\ and\ \citenamefont
		{Maley}}]{Houzet_FFLO2D_2002}%
	\BibitemOpen
	\bibfield  {author} {\bibinfo {author} {\bibfnamefont {M.}~\bibnamefont
			{Houzet}}, \bibinfo {author} {\bibfnamefont {A.}~\bibnamefont {Buzdin}},
		\bibinfo {author} {\bibfnamefont {L.}~\bibnamefont {Bulaevskii}}, \ and\
		\bibinfo {author} {\bibfnamefont {M.}~\bibnamefont {Maley}},\ }\bibfield
	{title} {\emph {\bibinfo {title} {{New Superconducting Phases in
					Field-Induced Organic Superconductor $\lambda$-(BETS)$_2$FeCl$_4$}},\ }}\href
	{\doibase 10.1103/PhysRevLett.88.227001} {\bibfield  {journal} {\bibinfo
			{journal} {Phys. Rev. Lett.}\ }\textbf {\bibinfo {volume} {88}},\ \bibinfo
		{pages} {227001} (\bibinfo {year} {2002})}\BibitemShut {NoStop}%
	\bibitem [{\citenamefont {Klemm}\ \emph {et~al.}(1975)\citenamefont {Klemm},
		\citenamefont {Luther},\ and\ \citenamefont {Beasley}}]{Klemm_Hc2_1975}%
	\BibitemOpen
	\bibfield  {author} {\bibinfo {author} {\bibfnamefont {R.~A.}\ \bibnamefont
			{Klemm}}, \bibinfo {author} {\bibfnamefont {A.}~\bibnamefont {Luther}}, \
		and\ \bibinfo {author} {\bibfnamefont {M.~R.}\ \bibnamefont {Beasley}},\
	}\bibfield  {title} {\emph {\bibinfo {title} {Theory of the upper critical
				field in layered superconductors},\ }}\href {\doibase
		10.1103/PhysRevB.12.877} {\bibfield  {journal} {\bibinfo  {journal} {Phys.
				Rev. B}\ }\textbf {\bibinfo {volume} {12}},\ \bibinfo {pages} {877} (\bibinfo
		{year} {1975})}\BibitemShut {NoStop}%
	\bibitem [{\citenamefont {Klemm}\ \emph {et~al.}(1974)\citenamefont {Klemm},
		\citenamefont {Beasley},\ and\ \citenamefont {Luther}}]{Klemm_Hc2_1974}%
	\BibitemOpen
	\bibfield  {author} {\bibinfo {author} {\bibfnamefont {R.~A.}\ \bibnamefont
			{Klemm}}, \bibinfo {author} {\bibfnamefont {M.~R.}\ \bibnamefont {Beasley}},
		\ and\ \bibinfo {author} {\bibfnamefont {A.}~\bibnamefont {Luther}},\
	}\bibfield  {title} {\emph {\bibinfo {title} {The upper critical field of
				layered superconductors},\ }}\href {\doibase 10.1007/BF00654905} {\bibfield
		{journal} {\bibinfo  {journal} {Journal of Low Temperature Physics}\ }\textbf
		{\bibinfo {volume} {16}},\ \bibinfo {pages} {607} (\bibinfo {year}
		{1974})}\BibitemShut {NoStop}%
	\bibitem [{\citenamefont {Barzykin}\ and\ \citenamefont
		{Gor'kov}(2002)}]{Barzykin_FFLO_SOC_2002}%
	\BibitemOpen
	\bibfield  {author} {\bibinfo {author} {\bibfnamefont {V.}~\bibnamefont
			{Barzykin}}\ and\ \bibinfo {author} {\bibfnamefont {L.~P.}\ \bibnamefont
			{Gor'kov}},\ }\bibfield  {title} {\emph {\bibinfo {title} {Inhomogeneous
				stripe phase revisited for surface superconductivity},\ }}\href {\doibase
		10.1103/PhysRevLett.89.227002} {\bibfield  {journal} {\bibinfo  {journal}
			{Phys. Rev. Lett.}\ }\textbf {\bibinfo {volume} {89}},\ \bibinfo {pages}
		{227002} (\bibinfo {year} {2002})}\BibitemShut {NoStop}%
	\bibitem [{\citenamefont {Zheng}\ \emph {et~al.}(2014)\citenamefont {Zheng},
		\citenamefont {Gong}, \citenamefont {Zhang}, \citenamefont {Zou},
		\citenamefont {Zhang},\ and\ \citenamefont {Guo}}]{Zheng_FFLO_SOC_2014}%
	\BibitemOpen
	\bibfield  {author} {\bibinfo {author} {\bibfnamefont {Z.}~\bibnamefont
			{Zheng}}, \bibinfo {author} {\bibfnamefont {M.}~\bibnamefont {Gong}},
		\bibinfo {author} {\bibfnamefont {Y.}~\bibnamefont {Zhang}}, \bibinfo
		{author} {\bibfnamefont {X.}~\bibnamefont {Zou}}, \bibinfo {author}
		{\bibfnamefont {C.}~\bibnamefont {Zhang}}, \ and\ \bibinfo {author}
		{\bibfnamefont {G.}~\bibnamefont {Guo}},\ }\bibfield  {title} {\emph
		{\bibinfo {title} {{FFLO Superfluids in 2D Spin-Orbit Coupled Fermi Gases}},\
	}}\href {\doibase 10.1038/srep06535} {\bibfield  {journal} {\bibinfo
			{journal} {Scientific Reports}\ }\textbf {\bibinfo {volume} {4}},\ \bibinfo
		{pages} {6535} (\bibinfo {year} {2014})}\BibitemShut {NoStop}%
	\bibitem [{\citenamefont {Yuan}\ and\ \citenamefont
		{Fu}(2021)}]{NYuan_FFLO_SOC_2021}%
	\BibitemOpen
	\bibfield  {author} {\bibinfo {author} {\bibfnamefont {N.~F.~Q.}\
			\bibnamefont {Yuan}}\ and\ \bibinfo {author} {\bibfnamefont {L.}~\bibnamefont
			{Fu}},\ }\bibfield  {title} {\emph {\bibinfo {title} {Topological metals and
				finite-momentum superconductors},\ }}\href {\doibase 10.1073/pnas.2019063118}
	{\bibfield  {journal} {\bibinfo  {journal} {Proceedings of the National
				Academy of Sciences}\ }\textbf {\bibinfo {volume} {118}},\ \bibinfo {pages}
		{e2019063118} (\bibinfo {year} {2021})}\BibitemShut {NoStop}%
	\bibitem [{\citenamefont {Xi}\ \emph {et~al.}(2016)\citenamefont {Xi},
		\citenamefont {Wang}, \citenamefont {Zhao}, \citenamefont {Park},
		\citenamefont {Law}, \citenamefont {Berger}, \citenamefont {Forró},
		\citenamefont {Shan},\ and\ \citenamefont {Mak}}]{XxXi_NbSe2_2016}%
	\BibitemOpen
	\bibfield  {author} {\bibinfo {author} {\bibfnamefont {X.}~\bibnamefont
			{Xi}}, \bibinfo {author} {\bibfnamefont {Z.}~\bibnamefont {Wang}}, \bibinfo
		{author} {\bibfnamefont {W.}~\bibnamefont {Zhao}}, \bibinfo {author}
		{\bibfnamefont {J.-H.}\ \bibnamefont {Park}}, \bibinfo {author}
		{\bibfnamefont {K.~T.}\ \bibnamefont {Law}}, \bibinfo {author} {\bibfnamefont
			{H.}~\bibnamefont {Berger}}, \bibinfo {author} {\bibfnamefont
			{L.}~\bibnamefont {Forró}}, \bibinfo {author} {\bibfnamefont
			{J.}~\bibnamefont {Shan}}, \ and\ \bibinfo {author} {\bibfnamefont {K.~F.}\
			\bibnamefont {Mak}},\ }\bibfield  {title} {\emph {\bibinfo {title} {{Ising
					pairing in superconducting NbSe$_2$ atomic layers}},\ }}\href {\doibase
		10.1038/nphys3538} {\bibfield  {journal} {\bibinfo  {journal} {Nature
				Physics}\ }\textbf {\bibinfo {volume} {12}},\ \bibinfo {pages} {139}
		(\bibinfo {year} {2016})}\BibitemShut {NoStop}%
	\bibitem [{\citenamefont {Saito}\ \emph {et~al.}(2016)\citenamefont {Saito},
		\citenamefont {Nakamura}, \citenamefont {Bahramy}, \citenamefont {Kohama},
		\citenamefont {Ye}, \citenamefont {Kasahara}, \citenamefont {Nakagawa},
		\citenamefont {Onga}, \citenamefont {Tokunaga}, \citenamefont {Nojima},
		\citenamefont {Yanase},\ and\ \citenamefont {Iwasa}}]{Saito_Ising_2016}%
	\BibitemOpen
	\bibfield  {author} {\bibinfo {author} {\bibfnamefont {Y.}~\bibnamefont
			{Saito}}, \bibinfo {author} {\bibfnamefont {Y.}~\bibnamefont {Nakamura}},
		\bibinfo {author} {\bibfnamefont {M.~S.}\ \bibnamefont {Bahramy}}, \bibinfo
		{author} {\bibfnamefont {Y.}~\bibnamefont {Kohama}}, \bibinfo {author}
		{\bibfnamefont {J.}~\bibnamefont {Ye}}, \bibinfo {author} {\bibfnamefont
			{Y.}~\bibnamefont {Kasahara}}, \bibinfo {author} {\bibfnamefont
			{Y.}~\bibnamefont {Nakagawa}}, \bibinfo {author} {\bibfnamefont
			{M.}~\bibnamefont {Onga}}, \bibinfo {author} {\bibfnamefont {M.}~\bibnamefont
			{Tokunaga}}, \bibinfo {author} {\bibfnamefont {T.}~\bibnamefont {Nojima}},
		\bibinfo {author} {\bibfnamefont {Y.}~\bibnamefont {Yanase}}, \ and\ \bibinfo
		{author} {\bibfnamefont {Y.}~\bibnamefont {Iwasa}},\ }\bibfield  {title}
	{\emph {\bibinfo {title} {{Superconductivity protected by spin–valley
					locking in ion-gated MoS$_2$}},\ }}\href {\doibase 10.1038/nphys3580}
	{\bibfield  {journal} {\bibinfo  {journal} {Nature Physics}\ }\textbf
		{\bibinfo {volume} {12}},\ \bibinfo {pages} {144} (\bibinfo {year}
		{2016})}\BibitemShut {NoStop}%
	\bibitem [{\citenamefont {Lu}\ \emph {et~al.}(2015)\citenamefont {Lu},
		\citenamefont {Zheliuk}, \citenamefont {Leermakers}, \citenamefont {Yuan},
		\citenamefont {Zeitler}, \citenamefont {Law},\ and\ \citenamefont
		{Ye}}]{Lu_Ising_2015}%
	\BibitemOpen
	\bibfield  {author} {\bibinfo {author} {\bibfnamefont {J.~M.}\ \bibnamefont
			{Lu}}, \bibinfo {author} {\bibfnamefont {O.}~\bibnamefont {Zheliuk}},
		\bibinfo {author} {\bibfnamefont {I.}~\bibnamefont {Leermakers}}, \bibinfo
		{author} {\bibfnamefont {N.~F.~Q.}\ \bibnamefont {Yuan}}, \bibinfo {author}
		{\bibfnamefont {U.}~\bibnamefont {Zeitler}}, \bibinfo {author} {\bibfnamefont
			{K.~T.}\ \bibnamefont {Law}}, \ and\ \bibinfo {author} {\bibfnamefont
			{J.~T.}\ \bibnamefont {Ye}},\ }\bibfield  {title} {\emph {\bibinfo {title}
			{{Evidence for two-dimensional Ising superconductivity in gated MoS$_2$}},\
	}}\href {\doibase 10.1126/science.aab2277} {\bibfield  {journal} {\bibinfo
			{journal} {Science}\ }\textbf {\bibinfo {volume} {350}},\ \bibinfo {pages}
		{1353} (\bibinfo {year} {2015})}\BibitemShut {NoStop}%
	\bibitem [{\citenamefont {Sohn}\ \emph {et~al.}(2018)\citenamefont {Sohn},
		\citenamefont {Xi}, \citenamefont {He}, \citenamefont {Jiang}, \citenamefont
		{Wang}, \citenamefont {Kang}, \citenamefont {Park}, \citenamefont {Berger},
		\citenamefont {Forró}, \citenamefont {Law}, \citenamefont {Shan},\ and\
		\citenamefont {Mak}}]{Sohn_Ising_2018}%
	\BibitemOpen
	\bibfield  {author} {\bibinfo {author} {\bibfnamefont {E.}~\bibnamefont
			{Sohn}}, \bibinfo {author} {\bibfnamefont {X.}~\bibnamefont {Xi}}, \bibinfo
		{author} {\bibfnamefont {W.-Y.}\ \bibnamefont {He}}, \bibinfo {author}
		{\bibfnamefont {S.}~\bibnamefont {Jiang}}, \bibinfo {author} {\bibfnamefont
			{Z.}~\bibnamefont {Wang}}, \bibinfo {author} {\bibfnamefont {K.}~\bibnamefont
			{Kang}}, \bibinfo {author} {\bibfnamefont {J.-H.}\ \bibnamefont {Park}},
		\bibinfo {author} {\bibfnamefont {H.}~\bibnamefont {Berger}}, \bibinfo
		{author} {\bibfnamefont {L.}~\bibnamefont {Forró}}, \bibinfo {author}
		{\bibfnamefont {K.~T.}\ \bibnamefont {Law}}, \bibinfo {author} {\bibfnamefont
			{J.}~\bibnamefont {Shan}}, \ and\ \bibinfo {author} {\bibfnamefont {K.~F.}\
			\bibnamefont {Mak}},\ }\bibfield  {title} {\emph {\bibinfo {title} {{An
					unusual continuous paramagnetic-limited superconducting phase transition in
					2D NbSe$_2$}},\ }}\href {\doibase 10.1038/s41563-018-0061-1} {\bibfield
		{journal} {\bibinfo  {journal} {Nature Materials}\ }\textbf {\bibinfo
			{volume} {17}},\ \bibinfo {pages} {504} (\bibinfo {year} {2018})}\BibitemShut
	{NoStop}%
	\bibitem [{\citenamefont {de~la Barrera}\ \emph {et~al.}(2018)\citenamefont
		{de~la Barrera}, \citenamefont {Sinko}, \citenamefont {Gopalan},
		\citenamefont {Sivadas}, \citenamefont {Seyler}, \citenamefont {Watanabe},
		\citenamefont {Taniguchi}, \citenamefont {Tsen}, \citenamefont {Xu},
		\citenamefont {Xiao},\ and\ \citenamefont {Hunt}}]{Barrera_Ising_2018}%
	\BibitemOpen
	\bibfield  {author} {\bibinfo {author} {\bibfnamefont {S.~C.}\ \bibnamefont
			{de~la Barrera}}, \bibinfo {author} {\bibfnamefont {M.~R.}\ \bibnamefont
			{Sinko}}, \bibinfo {author} {\bibfnamefont {D.~P.}\ \bibnamefont {Gopalan}},
		\bibinfo {author} {\bibfnamefont {N.}~\bibnamefont {Sivadas}}, \bibinfo
		{author} {\bibfnamefont {K.~L.}\ \bibnamefont {Seyler}}, \bibinfo {author}
		{\bibfnamefont {K.}~\bibnamefont {Watanabe}}, \bibinfo {author}
		{\bibfnamefont {T.}~\bibnamefont {Taniguchi}}, \bibinfo {author}
		{\bibfnamefont {A.~W.}\ \bibnamefont {Tsen}}, \bibinfo {author}
		{\bibfnamefont {X.}~\bibnamefont {Xu}}, \bibinfo {author} {\bibfnamefont
			{D.}~\bibnamefont {Xiao}}, \ and\ \bibinfo {author} {\bibfnamefont {B.~M.}\
			\bibnamefont {Hunt}},\ }\bibfield  {title} {\emph {\bibinfo {title} {{Tuning
					Ising superconductivity with layer and spin-orbit coupling in two-dimensional
					transition-metal dichalcogenides}},\ }}\href {\doibase
		10.1038/s41467-018-03888-4} {\bibfield  {journal} {\bibinfo  {journal}
			{Nature Communications}\ }\textbf {\bibinfo {volume} {9}},\ \bibinfo {pages}
		{1427} (\bibinfo {year} {2018})}\BibitemShut {NoStop}%
	\bibitem [{\citenamefont {Wu}\ \emph {et~al.}(2023{\natexlab{a}})\citenamefont
		{Wu}, \citenamefont {Wu},\ and\ \citenamefont {Yao}}]{YMWu_PDW_2023}%
	\BibitemOpen
	\bibfield  {author} {\bibinfo {author} {\bibfnamefont {Y.-M.}\ \bibnamefont
			{Wu}}, \bibinfo {author} {\bibfnamefont {Z.}~\bibnamefont {Wu}}, \ and\
		\bibinfo {author} {\bibfnamefont {H.}~\bibnamefont {Yao}},\ }\bibfield
	{title} {\emph {\bibinfo {title} {Pair-density-wave and chiral
				superconductivity in twisted bilayer transition metal dichalcogenides},\
	}}\href {\doibase 10.1103/PhysRevLett.130.126001} {\bibfield  {journal}
		{\bibinfo  {journal} {Phys. Rev. Lett.}\ }\textbf {\bibinfo {volume} {130}},\
		\bibinfo {pages} {126001} (\bibinfo {year} {2023}{\natexlab{a}})}\BibitemShut
	{NoStop}%
	\bibitem [{\citenamefont {Wu}\ \emph {et~al.}(2023{\natexlab{b}})\citenamefont
		{Wu}, \citenamefont {Wu},\ and\ \citenamefont {Wu}}]{YMWu_PDWvHs_2023}%
	\BibitemOpen
	\bibfield  {author} {\bibinfo {author} {\bibfnamefont {Z.}~\bibnamefont
			{Wu}}, \bibinfo {author} {\bibfnamefont {Y.-M.}\ \bibnamefont {Wu}}, \ and\
		\bibinfo {author} {\bibfnamefont {F.}~\bibnamefont {Wu}},\ }\bibfield
	{title} {\emph {\bibinfo {title} {Pair density wave and loop current promoted
				by van hove singularities in moir\'e systems},\ }}\href {\doibase
		10.1103/PhysRevB.107.045122} {\bibfield  {journal} {\bibinfo  {journal}
			{Phys. Rev. B}\ }\textbf {\bibinfo {volume} {107}},\ \bibinfo {pages}
		{045122} (\bibinfo {year} {2023}{\natexlab{b}})}\BibitemShut {NoStop}%
	\bibitem [{\citenamefont {Zhang}\ \emph {et~al.}(2024)\citenamefont {Zhang},
		\citenamefont {Wang}, \citenamefont {Liu}, \citenamefont {Fan}, \citenamefont
		{Cao},\ and\ \citenamefont {Xiao}}]{Zhang_TMDhomo_2024}%
	\BibitemOpen
	\bibfield  {author} {\bibinfo {author} {\bibfnamefont {X.-W.}\ \bibnamefont
			{Zhang}}, \bibinfo {author} {\bibfnamefont {C.}~\bibnamefont {Wang}},
		\bibinfo {author} {\bibfnamefont {X.}~\bibnamefont {Liu}}, \bibinfo {author}
		{\bibfnamefont {Y.}~\bibnamefont {Fan}}, \bibinfo {author} {\bibfnamefont
			{T.}~\bibnamefont {Cao}}, \ and\ \bibinfo {author} {\bibfnamefont
			{D.}~\bibnamefont {Xiao}},\ }\bibfield  {title} {\emph {\bibinfo {title}
			{{Polarization-driven band topology evolution in twisted MoTe$_2$ and
					WSe$_2$}},\ }}\href {\doibase 10.1038/s41467-024-48511-x} {\bibfield
		{journal} {\bibinfo  {journal} {Nature Communications}\ }\textbf {\bibinfo
			{volume} {15}},\ \bibinfo {pages} {4223} (\bibinfo {year}
		{2024})}\BibitemShut {NoStop}%
	\bibitem [{\citenamefont {Wu}\ \emph {et~al.}(2019)\citenamefont {Wu},
		\citenamefont {Lovorn}, \citenamefont {Tutuc}, \citenamefont {Martin},\ and\
		\citenamefont {MacDonald}}]{FWu_TMDhomoBi_2019}%
	\BibitemOpen
	\bibfield  {author} {\bibinfo {author} {\bibfnamefont {F.}~\bibnamefont
			{Wu}}, \bibinfo {author} {\bibfnamefont {T.}~\bibnamefont {Lovorn}}, \bibinfo
		{author} {\bibfnamefont {E.}~\bibnamefont {Tutuc}}, \bibinfo {author}
		{\bibfnamefont {I.}~\bibnamefont {Martin}}, \ and\ \bibinfo {author}
		{\bibfnamefont {A.~H.}\ \bibnamefont {MacDonald}},\ }\bibfield  {title}
	{\emph {\bibinfo {title} {Topological insulators in twisted transition metal
				dichalcogenide homobilayers},\ }}\href {\doibase
		10.1103/PhysRevLett.122.086402} {\bibfield  {journal} {\bibinfo  {journal}
			{Phys. Rev. Lett.}\ }\textbf {\bibinfo {volume} {122}},\ \bibinfo {pages}
		{086402} (\bibinfo {year} {2019})}\BibitemShut {NoStop}%
	\bibitem [{\citenamefont {Devakul}\ \emph {et~al.}(2021)\citenamefont
		{Devakul}, \citenamefont {Crépel}, \citenamefont {Zhang},\ and\
		\citenamefont {Fu}}]{TDevakul_DFT_2021}%
	\BibitemOpen
	\bibfield  {author} {\bibinfo {author} {\bibfnamefont {T.}~\bibnamefont
			{Devakul}}, \bibinfo {author} {\bibfnamefont {V.}~\bibnamefont {Crépel}},
		\bibinfo {author} {\bibfnamefont {Y.}~\bibnamefont {Zhang}}, \ and\ \bibinfo
		{author} {\bibfnamefont {L.}~\bibnamefont {Fu}},\ }\bibfield  {title} {\emph
		{\bibinfo {title} {Magic in twisted transition metal dichalcogenide
				bilayers},\ }}\href {\doibase 10.1038/s41467-021-27042-9} {\bibfield
		{journal} {\bibinfo  {journal} {Nature Communications}\ }\textbf {\bibinfo
			{volume} {12}},\ \bibinfo {pages} {6730} (\bibinfo {year}
		{2021})}\BibitemShut {NoStop}%
	\bibitem [{\citenamefont {Yu}\ \emph {et~al.}(2019)\citenamefont {Yu},
		\citenamefont {Chen},\ and\ \citenamefont {Yao}}]{HYu_TMDhomo_2019}%
	\BibitemOpen
	\bibfield  {author} {\bibinfo {author} {\bibfnamefont {H.}~\bibnamefont
			{Yu}}, \bibinfo {author} {\bibfnamefont {M.}~\bibnamefont {Chen}}, \ and\
		\bibinfo {author} {\bibfnamefont {W.}~\bibnamefont {Yao}},\ }\bibfield
	{title} {\emph {\bibinfo {title} {Giant magnetic field from moiré induced
				berry phase in homobilayer semiconductors},\ }}\href {\doibase
		10.1093/nsr/nwz117} {\bibfield  {journal} {\bibinfo  {journal} {National
				Science Review}\ }\textbf {\bibinfo {volume} {7}},\ \bibinfo {pages} {12}
		(\bibinfo {year} {2019})}\BibitemShut {NoStop}%
	\bibitem [{\citenamefont {Zhai}\ and\ \citenamefont
		{Yao}(2020)}]{DZhai_TMDhomo_2020}%
	\BibitemOpen
	\bibfield  {author} {\bibinfo {author} {\bibfnamefont {D.}~\bibnamefont
			{Zhai}}\ and\ \bibinfo {author} {\bibfnamefont {W.}~\bibnamefont {Yao}},\
	}\bibfield  {title} {\emph {\bibinfo {title} {Theory of tunable flux lattices
				in the homobilayer moir\'e of twisted and uniformly strained transition metal
				dichalcogenides},\ }}\href {\doibase 10.1103/PhysRevMaterials.4.094002}
	{\bibfield  {journal} {\bibinfo  {journal} {Phys. Rev. Mater.}\ }\textbf
		{\bibinfo {volume} {4}},\ \bibinfo {pages} {094002} (\bibinfo {year}
		{2020})}\BibitemShut {NoStop}%
	\bibitem [{\citenamefont {Naik}\ and\ \citenamefont
		{Jain}(2018)}]{Naik_TMDhomo_2018}%
	\BibitemOpen
	\bibfield  {author} {\bibinfo {author} {\bibfnamefont {M.~H.}\ \bibnamefont
			{Naik}}\ and\ \bibinfo {author} {\bibfnamefont {M.}~\bibnamefont {Jain}},\
	}\bibfield  {title} {\emph {\bibinfo {title} {Ultraflatbands and shear
				solitons in moir\'e patterns of twisted bilayer transition metal
				dichalcogenides},\ }}\href {\doibase 10.1103/PhysRevLett.121.266401}
	{\bibfield  {journal} {\bibinfo  {journal} {Phys. Rev. Lett.}\ }\textbf
		{\bibinfo {volume} {121}},\ \bibinfo {pages} {266401} (\bibinfo {year}
		{2018})}\BibitemShut {NoStop}%
	\bibitem [{\citenamefont {Kundu}\ \emph {et~al.}(2022)\citenamefont {Kundu},
		\citenamefont {Naik}, \citenamefont {Krishnamurthy},\ and\ \citenamefont
		{Jain}}]{Kundu_TMDhomo_2022}%
	\BibitemOpen
	\bibfield  {author} {\bibinfo {author} {\bibfnamefont {S.}~\bibnamefont
			{Kundu}}, \bibinfo {author} {\bibfnamefont {M.~H.}\ \bibnamefont {Naik}},
		\bibinfo {author} {\bibfnamefont {H.~R.}\ \bibnamefont {Krishnamurthy}}, \
		and\ \bibinfo {author} {\bibfnamefont {M.}~\bibnamefont {Jain}},\ }\bibfield
	{title} {\emph {\bibinfo {title} {{Moir\'e induced topology and flat bands in
					twisted bilayer WSe$_2$: A first-principles study}},\ }}\href {\doibase
		10.1103/PhysRevB.105.L081108} {\bibfield  {journal} {\bibinfo  {journal}
			{Phys. Rev. B}\ }\textbf {\bibinfo {volume} {105}},\ \bibinfo {pages}
		{L081108} (\bibinfo {year} {2022})}\BibitemShut {NoStop}%
	\bibitem [{\citenamefont {Hsu}\ \emph {et~al.}(2021)\citenamefont {Hsu},
		\citenamefont {Wu},\ and\ \citenamefont {Das~Sarma}}]{Hsu_VHS_2021}%
	\BibitemOpen
	\bibfield  {author} {\bibinfo {author} {\bibfnamefont {Y.-T.}\ \bibnamefont
			{Hsu}}, \bibinfo {author} {\bibfnamefont {F.}~\bibnamefont {Wu}}, \ and\
		\bibinfo {author} {\bibfnamefont {S.}~\bibnamefont {Das~Sarma}},\ }\bibfield
	{title} {\emph {\bibinfo {title} {Spin-valley locked instabilities in moir\'e
				transition metal dichalcogenides with conventional and higher-order van hove
				singularities},\ }}\href {\doibase 10.1103/PhysRevB.104.195134} {\bibfield
		{journal} {\bibinfo  {journal} {Phys. Rev. B}\ }\textbf {\bibinfo {volume}
			{104}},\ \bibinfo {pages} {195134} (\bibinfo {year} {2021})}\BibitemShut
	{NoStop}%
	\bibitem [{con()}]{continuum_paras}%
	\BibitemOpen
	\href@noop {} {}\bibinfo {note} {$\pmb{G}_j$ are the first-shell moir\'e
		reciprocal lattice vectors. We use the continuum model parameters from Ref.
		\cite{TDevakul_DFT_2021}: $V=9$ meV, $\psi=128^\circ$, $w=18$ meV, lattice
		constant $a_0 = 3.317 \text{ \AA}$, and the effective mass $m^*=0.43m_e$ with
		$m_e$ being the electron mass.}\BibitemShut {Stop}%
	\bibitem [{See()}]{SeeSM}%
	\BibitemOpen
	\href@noop {} {}\bibinfo {note} {See Supplemental Material (SM) at XXX for
		details of small-momentum and weak-field expansions, $\pmb{v}(\pmb{k})$ and
		$\pmb{M}(\pmb{k})$ in the MBZ, Fermi surface deformations in an in-plane
		field, phase diagrams for other twist angles from $2^\circ$-$5^\circ$ and in
		$\pmb{B}=B_y\hat{y}$.}\BibitemShut {Stop}%
	\bibitem [{\citenamefont {Wu}\ and\ \citenamefont
		{Das~Sarma}(2019)}]{FWu_TBG_Bpara_2019}%
	\BibitemOpen
	\bibfield  {author} {\bibinfo {author} {\bibfnamefont {F.}~\bibnamefont
			{Wu}}\ and\ \bibinfo {author} {\bibfnamefont {S.}~\bibnamefont {Das~Sarma}},\
	}\bibfield  {title} {\emph {\bibinfo {title} {Identification of
				superconducting pairing symmetry in twisted bilayer graphene using in-plane
				magnetic field and strain},\ }}\href {\doibase 10.1103/PhysRevB.99.220507}
	{\bibfield  {journal} {\bibinfo  {journal} {Phys. Rev. B}\ }\textbf {\bibinfo
			{volume} {99}},\ \bibinfo {pages} {220507} (\bibinfo {year}
		{2019})}\BibitemShut {NoStop}%
	\bibitem [{\citenamefont {Qin}\ and\ \citenamefont
		{MacDonald}(2021)}]{WQin_TTG_Bpara_2021}%
	\BibitemOpen
	\bibfield  {author} {\bibinfo {author} {\bibfnamefont {W.}~\bibnamefont
			{Qin}}\ and\ \bibinfo {author} {\bibfnamefont {A.~H.}\ \bibnamefont
			{MacDonald}},\ }\bibfield  {title} {\emph {\bibinfo {title} {In-plane
				critical magnetic fields in magic-angle twisted trilayer graphene},\ }}\href
	{\doibase 10.1103/PhysRevLett.127.097001} {\bibfield  {journal} {\bibinfo
			{journal} {Phys. Rev. Lett.}\ }\textbf {\bibinfo {volume} {127}},\ \bibinfo
		{pages} {097001} (\bibinfo {year} {2021})}\BibitemShut {NoStop}%
	\bibitem [{\citenamefont {Lake}\ and\ \citenamefont
		{Senthil}(2021)}]{ELake_TTG_SC_2021}%
	\BibitemOpen
	\bibfield  {author} {\bibinfo {author} {\bibfnamefont {E.}~\bibnamefont
			{Lake}}\ and\ \bibinfo {author} {\bibfnamefont {T.}~\bibnamefont {Senthil}},\
	}\bibfield  {title} {\emph {\bibinfo {title} {Reentrant superconductivity
				through a quantum lifshitz transition in twisted trilayer graphene},\ }}\href
	{\doibase 10.1103/PhysRevB.104.174505} {\bibfield  {journal} {\bibinfo
			{journal} {Phys. Rev. B}\ }\textbf {\bibinfo {volume} {104}},\ \bibinfo
		{pages} {174505} (\bibinfo {year} {2021})}\BibitemShut {NoStop}%
	\bibitem [{\citenamefont {Holleis}\ \emph {et~al.}(2025)\citenamefont
		{Holleis}, \citenamefont {Patterson}, \citenamefont {Zhang}, \citenamefont
		{Vituri}, \citenamefont {Yoo}, \citenamefont {Zhou}, \citenamefont
		{Taniguchi}, \citenamefont {Watanabe}, \citenamefont {Berg}, \citenamefont
		{Nadj-Perge},\ and\ \citenamefont {Young}}]{Holleis_BG_SC_2025}%
	\BibitemOpen
	\bibfield  {author} {\bibinfo {author} {\bibfnamefont {L.}~\bibnamefont
			{Holleis}}, \bibinfo {author} {\bibfnamefont {C.~L.}\ \bibnamefont
			{Patterson}}, \bibinfo {author} {\bibfnamefont {Y.}~\bibnamefont {Zhang}},
		\bibinfo {author} {\bibfnamefont {Y.}~\bibnamefont {Vituri}}, \bibinfo
		{author} {\bibfnamefont {H.~M.}\ \bibnamefont {Yoo}}, \bibinfo {author}
		{\bibfnamefont {H.}~\bibnamefont {Zhou}}, \bibinfo {author} {\bibfnamefont
			{T.}~\bibnamefont {Taniguchi}}, \bibinfo {author} {\bibfnamefont
			{K.}~\bibnamefont {Watanabe}}, \bibinfo {author} {\bibfnamefont
			{E.}~\bibnamefont {Berg}}, \bibinfo {author} {\bibfnamefont {S.}~\bibnamefont
			{Nadj-Perge}}, \ and\ \bibinfo {author} {\bibfnamefont {A.~F.}\ \bibnamefont
			{Young}},\ }\bibfield  {title} {\emph {\bibinfo {title} {Nematicity and
				orbital depairing in superconducting bernal bilayer graphene},\ }}\href
	{\doibase 10.1038/s41567-024-02776-7} {\bibfield  {journal} {\bibinfo
			{journal} {Nat. Phys.}\ }\textbf {\bibinfo {volume} {21}},\ \bibinfo {pages}
		{444} (\bibinfo {year} {2025})}\BibitemShut {NoStop}%
	\bibitem [{Note1()}]{Note1}%
	\BibitemOpen
	\bibinfo {note} {Due to the $C_{3z}$ rotational symmetry of tWSe$_2$, the
		critical in-plane field remains the same for equivalent
		directions.}\BibitemShut {Stop}%
	\bibitem [{\citenamefont {Shimahara}(1994)}]{Shimahara_FF_1994}%
	\BibitemOpen
	\bibfield  {author} {\bibinfo {author} {\bibfnamefont {H.}~\bibnamefont
			{Shimahara}},\ }\bibfield  {title} {\emph {\bibinfo {title} {{Fulde-Ferrell
					state in quasi-two-dimensional superconductors}},\ }}\href {\doibase
		10.1103/PhysRevB.50.12760} {\bibfield  {journal} {\bibinfo  {journal} {Phys.
				Rev. B}\ }\textbf {\bibinfo {volume} {50}},\ \bibinfo {pages} {12760}
		(\bibinfo {year} {1994})}\BibitemShut {NoStop}%
	\bibitem [{\citenamefont {Radzihovsky}(2011)}]{Radzihovsky_LO_2011}%
	\BibitemOpen
	\bibfield  {author} {\bibinfo {author} {\bibfnamefont {L.}~\bibnamefont
			{Radzihovsky}},\ }\bibfield  {title} {\emph {\bibinfo {title} {Fluctuations
				and phase transitions in larkin-ovchinnikov liquid-crystal states of a
				population-imbalanced resonant fermi gas},\ }}\href {\doibase
		10.1103/PhysRevA.84.023611} {\bibfield  {journal} {\bibinfo  {journal} {Phys.
				Rev. A}\ }\textbf {\bibinfo {volume} {84}},\ \bibinfo {pages} {023611}
		(\bibinfo {year} {2011})}\BibitemShut {NoStop}%
	\bibitem [{\citenamefont {Yang}(2001)}]{KYang_FFLO_2001}%
	\BibitemOpen
	\bibfield  {author} {\bibinfo {author} {\bibfnamefont {K.}~\bibnamefont
			{Yang}},\ }\bibfield  {title} {\emph {\bibinfo {title} {Inhomogeneous
				superconducting state in quasi-one-dimensional systems},\ }}\href {\doibase
		10.1103/PhysRevB.63.140511} {\bibfield  {journal} {\bibinfo  {journal} {Phys.
				Rev. B}\ }\textbf {\bibinfo {volume} {63}},\ \bibinfo {pages} {140511}
		(\bibinfo {year} {2001})}\BibitemShut {NoStop}%
	\bibitem [{Note2()}]{Note2}%
	\BibitemOpen
	\bibinfo {note} {The tricritical point temperature of $3.65^\circ $ tWSe$_2$,
		$T_{\protect \rm tric}/T_{c0} \approx 0.56$, coincidentally matches that of
		Ref.\cite {Casalbuoni_FFLO_2004}. However, in tWSe$_2$, this value depends on
		the interlayer coupling strength: as the interlayer coupling decreases (or
		the twist angle increases), the tricritical points shifts to a higher
		$T_{\protect \rm tric}/T_{c0}$ and lower $B/B_p$. Unlike in conventional
		Zeeman-driven FFLO phase diagram, where the transition from $q=0$ to
		$q\protect \neq 0$ superconductivity at $T=0$ occurs at the Pauli limit
		$B/B_p = 1$, in Ising superconductors, this transition takes place at a
		higher $B/B_p$ due to orbital effects and Ising SOC.}\BibitemShut {Stop}%
	\bibitem [{\citenamefont {Ando}\ \emph {et~al.}(2020)\citenamefont {Ando},
		\citenamefont {Miyasaka}, \citenamefont {Li}, \citenamefont {Ishizuka},
		\citenamefont {Arakawa}, \citenamefont {Shiota}, \citenamefont {Moriyama},
		\citenamefont {Yanase},\ and\ \citenamefont {Ono}}]{Ando_diode_2020}%
	\BibitemOpen
	\bibfield  {author} {\bibinfo {author} {\bibfnamefont {F.}~\bibnamefont
			{Ando}}, \bibinfo {author} {\bibfnamefont {Y.}~\bibnamefont {Miyasaka}},
		\bibinfo {author} {\bibfnamefont {T.}~\bibnamefont {Li}}, \bibinfo {author}
		{\bibfnamefont {J.}~\bibnamefont {Ishizuka}}, \bibinfo {author}
		{\bibfnamefont {T.}~\bibnamefont {Arakawa}}, \bibinfo {author} {\bibfnamefont
			{Y.}~\bibnamefont {Shiota}}, \bibinfo {author} {\bibfnamefont
			{T.}~\bibnamefont {Moriyama}}, \bibinfo {author} {\bibfnamefont
			{Y.}~\bibnamefont {Yanase}}, \ and\ \bibinfo {author} {\bibfnamefont
			{T.}~\bibnamefont {Ono}},\ }\bibfield  {title} {\emph {\bibinfo {title}
			{Observation of superconducting diode effect},\ }}\href {\doibase
		10.1038/s41586-020-2590-4} {\bibfield  {journal} {\bibinfo  {journal}
			{Nature}\ }\textbf {\bibinfo {volume} {584}},\ \bibinfo {pages} {373}
		(\bibinfo {year} {2020})}\BibitemShut {NoStop}%
	\bibitem [{\citenamefont {Bauriedl}\ \emph {et~al.}(2022)\citenamefont
		{Bauriedl}, \citenamefont {Bäuml}, \citenamefont {Fuchs}, \citenamefont
		{Baumgartner}, \citenamefont {Paulik}, \citenamefont {Bauer}, \citenamefont
		{Lin}, \citenamefont {Lupton}, \citenamefont {Taniguchi}, \citenamefont
		{Watanabe}, \citenamefont {Strunk},\ and\ \citenamefont
		{Paradiso}}]{Bauriedl_diode_2022}%
	\BibitemOpen
	\bibfield  {author} {\bibinfo {author} {\bibfnamefont {L.}~\bibnamefont
			{Bauriedl}}, \bibinfo {author} {\bibfnamefont {C.}~\bibnamefont {Bäuml}},
		\bibinfo {author} {\bibfnamefont {L.}~\bibnamefont {Fuchs}}, \bibinfo
		{author} {\bibfnamefont {C.}~\bibnamefont {Baumgartner}}, \bibinfo {author}
		{\bibfnamefont {N.}~\bibnamefont {Paulik}}, \bibinfo {author} {\bibfnamefont
			{J.~M.}\ \bibnamefont {Bauer}}, \bibinfo {author} {\bibfnamefont {K.-Q.}\
			\bibnamefont {Lin}}, \bibinfo {author} {\bibfnamefont {J.~M.}\ \bibnamefont
			{Lupton}}, \bibinfo {author} {\bibfnamefont {T.}~\bibnamefont {Taniguchi}},
		\bibinfo {author} {\bibfnamefont {K.}~\bibnamefont {Watanabe}}, \bibinfo
		{author} {\bibfnamefont {C.}~\bibnamefont {Strunk}}, \ and\ \bibinfo {author}
		{\bibfnamefont {N.}~\bibnamefont {Paradiso}},\ }\bibfield  {title} {\emph
		{\bibinfo {title} {{Supercurrent diode effect and magnetochiral anisotropy in
					few-layer NbSe$_2$}},\ }}\href {\doibase 10.1038/s41467-022-31954-5}
	{\bibfield  {journal} {\bibinfo  {journal} {Nature Communications}\ }\textbf
		{\bibinfo {volume} {13}},\ \bibinfo {pages} {4266} (\bibinfo {year}
		{2022})}\BibitemShut {NoStop}%
	\bibitem [{\citenamefont {Baumgartner}\ \emph {et~al.}(2022)\citenamefont
		{Baumgartner}, \citenamefont {Fuchs}, \citenamefont {Costa}, \citenamefont
		{Reinhardt}, \citenamefont {Gronin}, \citenamefont {Gardner}, \citenamefont
		{Lindemann}, \citenamefont {Manfra}, \citenamefont {Faria~Junior},
		\citenamefont {Kochan}, \citenamefont {Fabian}, \citenamefont {Paradiso},\
		and\ \citenamefont {Strunk}}]{Baumgartner_diode_2022}%
	\BibitemOpen
	\bibfield  {author} {\bibinfo {author} {\bibfnamefont {C.}~\bibnamefont
			{Baumgartner}}, \bibinfo {author} {\bibfnamefont {L.}~\bibnamefont {Fuchs}},
		\bibinfo {author} {\bibfnamefont {A.}~\bibnamefont {Costa}}, \bibinfo
		{author} {\bibfnamefont {S.}~\bibnamefont {Reinhardt}}, \bibinfo {author}
		{\bibfnamefont {S.}~\bibnamefont {Gronin}}, \bibinfo {author} {\bibfnamefont
			{G.~C.}\ \bibnamefont {Gardner}}, \bibinfo {author} {\bibfnamefont
			{T.}~\bibnamefont {Lindemann}}, \bibinfo {author} {\bibfnamefont {M.~J.}\
			\bibnamefont {Manfra}}, \bibinfo {author} {\bibfnamefont {P.~E.}\
			\bibnamefont {Faria~Junior}}, \bibinfo {author} {\bibfnamefont
			{D.}~\bibnamefont {Kochan}}, \bibinfo {author} {\bibfnamefont
			{J.}~\bibnamefont {Fabian}}, \bibinfo {author} {\bibfnamefont
			{N.}~\bibnamefont {Paradiso}}, \ and\ \bibinfo {author} {\bibfnamefont
			{C.}~\bibnamefont {Strunk}},\ }\bibfield  {title} {\emph {\bibinfo {title}
			{Supercurrent rectification and magnetochiral effects in symmetric josephson
				junctions},\ }}\href {\doibase 10.1038/s41565-021-01009-9} {\bibfield
		{journal} {\bibinfo  {journal} {Nature Nanotechnology}\ }\textbf {\bibinfo
			{volume} {17}},\ \bibinfo {pages} {39} (\bibinfo {year} {2022})}\BibitemShut
	{NoStop}%
	\bibitem [{\citenamefont {Lin}\ \emph {et~al.}(2022)\citenamefont {Lin},
		\citenamefont {Siriviboon}, \citenamefont {Scammell}, \citenamefont {Liu},
		\citenamefont {Rhodes}, \citenamefont {Watanabe}, \citenamefont {Taniguchi},
		\citenamefont {Hone}, \citenamefont {Scheurer},\ and\ \citenamefont
		{Li}}]{Lin_diode_2022}%
	\BibitemOpen
	\bibfield  {author} {\bibinfo {author} {\bibfnamefont {J.-X.}\ \bibnamefont
			{Lin}}, \bibinfo {author} {\bibfnamefont {P.}~\bibnamefont {Siriviboon}},
		\bibinfo {author} {\bibfnamefont {H.~D.}\ \bibnamefont {Scammell}}, \bibinfo
		{author} {\bibfnamefont {S.}~\bibnamefont {Liu}}, \bibinfo {author}
		{\bibfnamefont {D.}~\bibnamefont {Rhodes}}, \bibinfo {author} {\bibfnamefont
			{K.}~\bibnamefont {Watanabe}}, \bibinfo {author} {\bibfnamefont
			{T.}~\bibnamefont {Taniguchi}}, \bibinfo {author} {\bibfnamefont
			{J.}~\bibnamefont {Hone}}, \bibinfo {author} {\bibfnamefont {M.~S.}\
			\bibnamefont {Scheurer}}, \ and\ \bibinfo {author} {\bibfnamefont
			{J.}~\bibnamefont {Li}},\ }\bibfield  {title} {\emph {\bibinfo {title}
			{Zero-field superconducting diode effect in small-twist-angle trilayer
				graphene},\ }}\href {\doibase 10.1038/s41567-022-01700-1} {\bibfield
		{journal} {\bibinfo  {journal} {Nature Physics}\ }\textbf {\bibinfo {volume}
			{18}},\ \bibinfo {pages} {1221} (\bibinfo {year} {2022})}\BibitemShut
	{NoStop}%
	\bibitem [{\citenamefont {Pal}\ \emph {et~al.}(2022)\citenamefont {Pal},
		\citenamefont {Chakraborty}, \citenamefont {Sivakumar}, \citenamefont
		{Davydova}, \citenamefont {Gopi}, \citenamefont {Pandeya}, \citenamefont
		{Krieger}, \citenamefont {Zhang}, \citenamefont {Date}, \citenamefont {Ju},
		\citenamefont {Yuan}, \citenamefont {Schröter}, \citenamefont {Fu},\ and\
		\citenamefont {Parkin}}]{Pal_diode_2022}%
	\BibitemOpen
	\bibfield  {author} {\bibinfo {author} {\bibfnamefont {B.}~\bibnamefont
			{Pal}}, \bibinfo {author} {\bibfnamefont {A.}~\bibnamefont {Chakraborty}},
		\bibinfo {author} {\bibfnamefont {P.~K.}\ \bibnamefont {Sivakumar}}, \bibinfo
		{author} {\bibfnamefont {M.}~\bibnamefont {Davydova}}, \bibinfo {author}
		{\bibfnamefont {A.~K.}\ \bibnamefont {Gopi}}, \bibinfo {author}
		{\bibfnamefont {A.~K.}\ \bibnamefont {Pandeya}}, \bibinfo {author}
		{\bibfnamefont {J.~A.}\ \bibnamefont {Krieger}}, \bibinfo {author}
		{\bibfnamefont {Y.}~\bibnamefont {Zhang}}, \bibinfo {author} {\bibfnamefont
			{M.}~\bibnamefont {Date}}, \bibinfo {author} {\bibfnamefont {S.}~\bibnamefont
			{Ju}}, \bibinfo {author} {\bibfnamefont {N.}~\bibnamefont {Yuan}}, \bibinfo
		{author} {\bibfnamefont {N.~B.~M.}\ \bibnamefont {Schröter}}, \bibinfo
		{author} {\bibfnamefont {L.}~\bibnamefont {Fu}}, \ and\ \bibinfo {author}
		{\bibfnamefont {S.~S.~P.}\ \bibnamefont {Parkin}},\ }\bibfield  {title}
	{\emph {\bibinfo {title} {Josephson diode effect from cooper pair momentum in
				a topological semimetal},\ }}\href {\doibase 10.1038/s41567-022-01699-5}
	{\bibfield  {journal} {\bibinfo  {journal} {Nature Physics}\ }\textbf
		{\bibinfo {volume} {18}},\ \bibinfo {pages} {1228} (\bibinfo {year}
		{2022})}\BibitemShut {NoStop}%
	\bibitem [{\citenamefont {Yuan}\ and\ \citenamefont
		{Fu}(2022)}]{NYuan_diode_2022}%
	\BibitemOpen
	\bibfield  {author} {\bibinfo {author} {\bibfnamefont {N.~F.~Q.}\
			\bibnamefont {Yuan}}\ and\ \bibinfo {author} {\bibfnamefont {L.}~\bibnamefont
			{Fu}},\ }\bibfield  {title} {\emph {\bibinfo {title} {Supercurrent diode
				effect and finite-momentum superconductors},\ }}\href {\doibase
		10.1073/pnas.2119548119} {\bibfield  {journal} {\bibinfo  {journal}
			{Proceedings of the National Academy of Sciences}\ }\textbf {\bibinfo
			{volume} {119}},\ \bibinfo {pages} {e2119548119} (\bibinfo {year}
		{2022})}\BibitemShut {NoStop}%
	\bibitem [{\citenamefont {Nakamura}\ \emph {et~al.}(2024)\citenamefont
		{Nakamura}, \citenamefont {Daido},\ and\ \citenamefont
		{Yanase}}]{Nakamura_diode_2024}%
	\BibitemOpen
	\bibfield  {author} {\bibinfo {author} {\bibfnamefont {K.}~\bibnamefont
			{Nakamura}}, \bibinfo {author} {\bibfnamefont {A.}~\bibnamefont {Daido}}, \
		and\ \bibinfo {author} {\bibfnamefont {Y.}~\bibnamefont {Yanase}},\
	}\bibfield  {title} {\emph {\bibinfo {title} {Orbital effect on the intrinsic
				superconducting diode effect},\ }}\href {\doibase
		10.1103/PhysRevB.109.094501} {\bibfield  {journal} {\bibinfo  {journal}
			{Phys. Rev. B}\ }\textbf {\bibinfo {volume} {109}},\ \bibinfo {pages}
		{094501} (\bibinfo {year} {2024})}\BibitemShut {NoStop}%
	\bibitem [{\citenamefont {He}\ \emph {et~al.}(2022)\citenamefont {He},
		\citenamefont {Tanaka},\ and\ \citenamefont {Nagaosa}}]{He_diode_2022}%
	\BibitemOpen
	\bibfield  {author} {\bibinfo {author} {\bibfnamefont {J.~J.}\ \bibnamefont
			{He}}, \bibinfo {author} {\bibfnamefont {Y.}~\bibnamefont {Tanaka}}, \ and\
		\bibinfo {author} {\bibfnamefont {N.}~\bibnamefont {Nagaosa}},\ }\bibfield
	{title} {\emph {\bibinfo {title} {A phenomenological theory of superconductor
				diodes},\ }}\href {\doibase 10.1088/1367-2630/ac6766} {\bibfield  {journal}
		{\bibinfo  {journal} {New Journal of Physics}\ }\textbf {\bibinfo {volume}
			{24}},\ \bibinfo {pages} {053014} (\bibinfo {year} {2022})}\BibitemShut
	{NoStop}%
	\bibitem [{\citenamefont {Bankier}\ \emph {et~al.}(2025)\citenamefont
		{Bankier}, \citenamefont {Attias}, \citenamefont {Levchenko},\ and\
		\citenamefont {Khodas}}]{Bankier_diode_2025}%
	\BibitemOpen
	\bibfield  {author} {\bibinfo {author} {\bibfnamefont {I.}~\bibnamefont
			{Bankier}}, \bibinfo {author} {\bibfnamefont {L.}~\bibnamefont {Attias}},
		\bibinfo {author} {\bibfnamefont {A.}~\bibnamefont {Levchenko}}, \ and\
		\bibinfo {author} {\bibfnamefont {M.}~\bibnamefont {Khodas}},\ }\bibfield
	{title} {\emph {\bibinfo {title} {Superconducting diode effect in ising
				superconductors},\ }}\href {\doibase 10.1103/PhysRevB.111.L180505} {\bibfield
		{journal} {\bibinfo  {journal} {Phys. Rev. B}\ }\textbf {\bibinfo {volume}
			{111}},\ \bibinfo {pages} {L180505} (\bibinfo {year} {2025})}\BibitemShut
	{NoStop}%
	\bibitem [{\citenamefont {Pan}\ \emph {et~al.}(2020)\citenamefont {Pan},
		\citenamefont {Wu},\ and\ \citenamefont {Das~Sarma}}]{PanH2020}%
	\BibitemOpen
	\bibfield  {author} {\bibinfo {author} {\bibfnamefont {H.}~\bibnamefont
			{Pan}}, \bibinfo {author} {\bibfnamefont {F.}~\bibnamefont {Wu}}, \ and\
		\bibinfo {author} {\bibfnamefont {S.}~\bibnamefont {Das~Sarma}},\ }\bibfield
	{title} {\emph {\bibinfo {title} {Band topology, {{Hubbard}} model,
				{{Heisenberg}} model, and {{Dzyaloshinskii-Moriya}} interaction in twisted
				bilayer {{WSe}}$_2$},\ }}\href {\doibase 10.1103/PhysRevResearch.2.033087}
	{\bibfield  {journal} {\bibinfo  {journal} {Phys. Rev. Res.}\ }\textbf
		{\bibinfo {volume} {2}},\ \bibinfo {pages} {033087} (\bibinfo {year}
		{2020})}\BibitemShut {NoStop}%
	\bibitem [{\citenamefont {Han}\ \emph {et~al.}(2025)\citenamefont {Han},
		\citenamefont {Lu}, \citenamefont {Hadjri}, \citenamefont {Shi},
		\citenamefont {Wu}, \citenamefont {Xu}, \citenamefont {Yao}, \citenamefont
		{Cotten}, \citenamefont {Sedeh}, \citenamefont {Weldeyesus}, \citenamefont
		{Yang}, \citenamefont {Seo}, \citenamefont {Ye}, \citenamefont {Zhou},
		\citenamefont {Liu}, \citenamefont {Shi}, \citenamefont {Hua}, \citenamefont
		{Watanabe}, \citenamefont {Taniguchi}, \citenamefont {Xiong}, \citenamefont
		{Zumbühl}, \citenamefont {Fu},\ and\ \citenamefont {Ju}}]{LongJu_chiralSC}%
	\BibitemOpen
	\bibfield  {author} {\bibinfo {author} {\bibfnamefont {T.}~\bibnamefont
			{Han}}, \bibinfo {author} {\bibfnamefont {Z.}~\bibnamefont {Lu}}, \bibinfo
		{author} {\bibfnamefont {Z.}~\bibnamefont {Hadjri}}, \bibinfo {author}
		{\bibfnamefont {L.}~\bibnamefont {Shi}}, \bibinfo {author} {\bibfnamefont
			{Z.}~\bibnamefont {Wu}}, \bibinfo {author} {\bibfnamefont {W.}~\bibnamefont
			{Xu}}, \bibinfo {author} {\bibfnamefont {Y.}~\bibnamefont {Yao}}, \bibinfo
		{author} {\bibfnamefont {A.~A.}\ \bibnamefont {Cotten}}, \bibinfo {author}
		{\bibfnamefont {O.~S.}\ \bibnamefont {Sedeh}}, \bibinfo {author}
		{\bibfnamefont {H.}~\bibnamefont {Weldeyesus}}, \bibinfo {author}
		{\bibfnamefont {J.}~\bibnamefont {Yang}}, \bibinfo {author} {\bibfnamefont
			{J.}~\bibnamefont {Seo}}, \bibinfo {author} {\bibfnamefont {S.}~\bibnamefont
			{Ye}}, \bibinfo {author} {\bibfnamefont {M.}~\bibnamefont {Zhou}}, \bibinfo
		{author} {\bibfnamefont {H.}~\bibnamefont {Liu}}, \bibinfo {author}
		{\bibfnamefont {G.}~\bibnamefont {Shi}}, \bibinfo {author} {\bibfnamefont
			{Z.}~\bibnamefont {Hua}}, \bibinfo {author} {\bibfnamefont {K.}~\bibnamefont
			{Watanabe}}, \bibinfo {author} {\bibfnamefont {T.}~\bibnamefont {Taniguchi}},
		\bibinfo {author} {\bibfnamefont {P.}~\bibnamefont {Xiong}}, \bibinfo
		{author} {\bibfnamefont {D.~M.}\ \bibnamefont {Zumbühl}}, \bibinfo {author}
		{\bibfnamefont {L.}~\bibnamefont {Fu}}, \ and\ \bibinfo {author}
		{\bibfnamefont {L.}~\bibnamefont {Ju}},\ }\bibfield  {title} {\emph {\bibinfo
			{title} {Signatures of chiral superconductivity in rhombohedral graphene},\
	}}\href {\doibase 10.1038/s41586-025-09169-7} {\bibfield  {journal} {\bibinfo
			{journal} {Nature}\ }\textbf {\bibinfo {volume} {643}},\ \bibinfo {pages}
		{654} (\bibinfo {year} {2025})}\BibitemShut {NoStop}%
	\bibitem [{\citenamefont {Chou}\ \emph {et~al.}(2025)\citenamefont {Chou},
		\citenamefont {Zhu},\ and\ \citenamefont {Das~Sarma}}]{YChou_chiralSC_2024}%
	\BibitemOpen
	\bibfield  {author} {\bibinfo {author} {\bibfnamefont {Y.-Z.}\ \bibnamefont
			{Chou}}, \bibinfo {author} {\bibfnamefont {J.}~\bibnamefont {Zhu}}, \ and\
		\bibinfo {author} {\bibfnamefont {S.}~\bibnamefont {Das~Sarma}},\ }\bibfield
	{title} {\emph {\bibinfo {title} {Intravalley spin-polarized
				superconductivity in rhombohedral tetralayer graphene},\ }}\href {\doibase
		10.1103/PhysRevB.111.174523} {\bibfield  {journal} {\bibinfo  {journal}
			{Phys. Rev. B}\ }\textbf {\bibinfo {volume} {111}},\ \bibinfo {pages}
		{174523} (\bibinfo {year} {2025})}\BibitemShut {NoStop}%
	\bibitem [{\citenamefont {Yoon}\ \emph {et~al.}()\citenamefont {Yoon},
		\citenamefont {Xu}, \citenamefont {Barlas},\ and\ \citenamefont
		{Zhang}}]{Yoon_QM_SC_2025}%
	\BibitemOpen
	\bibfield  {author} {\bibinfo {author} {\bibfnamefont {C.}~\bibnamefont
			{Yoon}}, \bibinfo {author} {\bibfnamefont {T.}~\bibnamefont {Xu}}, \bibinfo
		{author} {\bibfnamefont {Y.}~\bibnamefont {Barlas}}, \ and\ \bibinfo {author}
		{\bibfnamefont {F.}~\bibnamefont {Zhang}},\ }\bibfield  {title} {\emph
		{\bibinfo {title} {{Quarter Metal Superconductivity}},\ }}\href
	{https://doi.org/10.48550/arXiv.2502.17555} {\ }\Eprint
	{http://arxiv.org/abs/arXiv:2502.17555}{arXiv:2502.17555}\BibitemShut
	{NoStop}%
	\bibitem [{\citenamefont {Kim}\ \emph {et~al.}(2025{\natexlab{b}})\citenamefont
		{Kim}, \citenamefont {Timmel}, \citenamefont {Ju},\ and\ \citenamefont
		{Wen}}]{Kim_chiralSC_2024}%
	\BibitemOpen
	\bibfield  {author} {\bibinfo {author} {\bibfnamefont {M.}~\bibnamefont
			{Kim}}, \bibinfo {author} {\bibfnamefont {A.}~\bibnamefont {Timmel}},
		\bibinfo {author} {\bibfnamefont {L.}~\bibnamefont {Ju}}, \ and\ \bibinfo
		{author} {\bibfnamefont {X.-G.}\ \bibnamefont {Wen}},\ }\bibfield  {title}
	{\emph {\bibinfo {title} {Topological chiral superconductivity beyond pairing
				in a fermi liquid},\ }}\href {\doibase 10.1103/PhysRevB.111.014508}
	{\bibfield  {journal} {\bibinfo  {journal} {Phys. Rev. B}\ }\textbf {\bibinfo
			{volume} {111}},\ \bibinfo {pages} {014508} (\bibinfo {year}
		{2025}{\natexlab{b}})}\BibitemShut {NoStop}%
	\bibitem [{\citenamefont {Qin}\ and\ \citenamefont {Wu}()}]{Qin_chiralSC_2025}%
	\BibitemOpen
	\bibfield  {author} {\bibinfo {author} {\bibfnamefont {Q.}~\bibnamefont
			{Qin}}\ and\ \bibinfo {author} {\bibfnamefont {C.}~\bibnamefont {Wu}},\
	}\bibfield  {title} {\emph {\bibinfo {title} {{Chiral finite-momentum
					superconductivity in the tetralayer graphene}},\ }}\href
	{https://doi.org/10.48550/arXiv.2412.07145} {\ }\Eprint
	{http://arxiv.org/abs/arXiv:2412.07145}{arXiv:2412.07145}\BibitemShut
	{NoStop}%
	\bibitem [{\citenamefont {Yang}\ and\ \citenamefont
		{Zhang}()}]{HYang_FFLO_2024}%
	\BibitemOpen
	\bibfield  {author} {\bibinfo {author} {\bibfnamefont {H.}~\bibnamefont
			{Yang}}\ and\ \bibinfo {author} {\bibfnamefont {Y.-H.}\ \bibnamefont
			{Zhang}},\ }\bibfield  {title} {\emph {\bibinfo {title} {{Topological
					incommensurate Fulde-Ferrell-Larkin-Ovchinnikov superconductor and Bogoliubov
					Fermi surface in rhombohedral tetralayer graphene}},\ }}\href
	{https://doi.org/10.48550/arXiv.2411.02503} {\ }\Eprint
	{http://arxiv.org/abs/arXiv:2411.02503}{arXiv:2411.02503}\BibitemShut
	{NoStop}%
	\bibitem [{\citenamefont {Ammar}\ and\ \citenamefont
		{Lin}()}]{Jahin_chiralSC_2024}%
	\BibitemOpen
	\bibfield  {author} {\bibinfo {author} {\bibfnamefont {J.}~\bibnamefont
			{Ammar}}\ and\ \bibinfo {author} {\bibfnamefont {S.-Z.}\ \bibnamefont
			{Lin}},\ }\bibfield  {title} {\emph {\bibinfo {title} {{Enhanced
					Kohn-Luttinger topological superconductivity in bands with nontrivial
					geometry}},\ }}\href {https://doi.org/10.48550/arXiv.2411.09664} {\ }\Eprint
	{http://arxiv.org/abs/arXiv:2411.09664}{arXiv:2411.09664}\BibitemShut
	{NoStop}%
	\bibitem [{\citenamefont {Sau}\ and\ \citenamefont {Wang}()}]{Sau_vortex_2024}%
	\BibitemOpen
	\bibfield  {author} {\bibinfo {author} {\bibfnamefont {J.~D.}\ \bibnamefont
			{Sau}}\ and\ \bibinfo {author} {\bibfnamefont {S.}~\bibnamefont {Wang}},\
	}\bibfield  {title} {\emph {\bibinfo {title} {{Theory of anomalous Hall
					effect from screened vortex charge in a phase disordered superconductor}},\
	}}\href {https://doi.org/10.48550/arXiv.2411.08969} {\ }\Eprint
	{http://arxiv.org/abs/arXiv:2411.08969}{arXiv:2411.08969}\BibitemShut
	{NoStop}%
	\bibitem [{\citenamefont {Wang}\ \emph {et~al.}()\citenamefont {Wang},
		\citenamefont {Gao},\ and\ \citenamefont {Yang}}]{Wang_chiralSC_2024}%
	\BibitemOpen
	\bibfield  {author} {\bibinfo {author} {\bibfnamefont {Y.-Q.}\ \bibnamefont
			{Wang}}, \bibinfo {author} {\bibfnamefont {Z.-Q.}\ \bibnamefont {Gao}}, \
		and\ \bibinfo {author} {\bibfnamefont {H.}~\bibnamefont {Yang}},\ }\bibfield
	{title} {\emph {\bibinfo {title} {{Chiral superconductivity from parent Chern
					band and its non-Abelian generalization}},\ }}\href
	{https://doi.org/10.48550/arXiv.2410.05384} {\ }\Eprint
	{http://arxiv.org/abs/arXiv:2410.05384}{arXiv:2410.05384}\BibitemShut
	{NoStop}%
	\bibitem [{\citenamefont {Geier}\ \emph {et~al.}()\citenamefont {Geier},
		\citenamefont {Davydova},\ and\ \citenamefont {Fu}}]{Geier_chiralSC_2024}%
	\BibitemOpen
	\bibfield  {author} {\bibinfo {author} {\bibfnamefont {M.}~\bibnamefont
			{Geier}}, \bibinfo {author} {\bibfnamefont {M.}~\bibnamefont {Davydova}}, \
		and\ \bibinfo {author} {\bibfnamefont {L.}~\bibnamefont {Fu}},\ }\bibfield
	{title} {\emph {\bibinfo {title} {{Chiral and topological superconductivity
					in isospin polarized multilayer graphene}},\ }}\href
	{https://doi.org/10.48550/arXiv.2409.13829} {\ }\Eprint
	{http://arxiv.org/abs/arXiv:2409.13829}{arXiv:2409.13829}\BibitemShut
	{NoStop}%
	\bibitem [{\citenamefont {Parra-Martinez}\ \emph {et~al.}()\citenamefont
		{Parra-Martinez}, \citenamefont {Jimeno-Pozo}, \citenamefont {Phong},
		\citenamefont {Sainz-Cruz}, \citenamefont {Kaplan}, \citenamefont {Emanuel},
		\citenamefont {Oreg}, \citenamefont {Pantaleon}, \citenamefont
		{Silva-Guillen},\ and\ \citenamefont {Guinea}}]{Guinea_chiralSC_2025}%
	\BibitemOpen
	\bibfield  {author} {\bibinfo {author} {\bibfnamefont {G.}~\bibnamefont
			{Parra-Martinez}}, \bibinfo {author} {\bibfnamefont {A.}~\bibnamefont
			{Jimeno-Pozo}}, \bibinfo {author} {\bibfnamefont {V.~T.}\ \bibnamefont
			{Phong}}, \bibinfo {author} {\bibfnamefont {H.}~\bibnamefont {Sainz-Cruz}},
		\bibinfo {author} {\bibfnamefont {D.}~\bibnamefont {Kaplan}}, \bibinfo
		{author} {\bibfnamefont {P.}~\bibnamefont {Emanuel}}, \bibinfo {author}
		{\bibfnamefont {Y.}~\bibnamefont {Oreg}}, \bibinfo {author} {\bibfnamefont
			{P.~A.}\ \bibnamefont {Pantaleon}}, \bibinfo {author} {\bibfnamefont {J.~A.}\
			\bibnamefont {Silva-Guillen}}, \ and\ \bibinfo {author} {\bibfnamefont
			{F.}~\bibnamefont {Guinea}},\ }\bibfield  {title} {\emph {\bibinfo {title}
			{{Band Renormalization, Quarter Metals, and Chiral Superconductivity in
					Rhombohedral Tetralayer Graphene}},\ }}\href
	{https://doi.org/10.48550/arXiv.2502.19474} {\ }\Eprint
	{http://arxiv.org/abs/arXiv:2502.19474}{arXiv:2502.19474}\BibitemShut
	{NoStop}%
	\bibitem [{\citenamefont {May-Mann}\ \emph {et~al.}()\citenamefont {May-Mann},
		\citenamefont {Helbig},\ and\ \citenamefont
		{Devakul}}]{MayMann_chiralSC_2025}%
	\BibitemOpen
	\bibfield  {author} {\bibinfo {author} {\bibfnamefont {J.}~\bibnamefont
			{May-Mann}}, \bibinfo {author} {\bibfnamefont {T.}~\bibnamefont {Helbig}}, \
		and\ \bibinfo {author} {\bibfnamefont {T.}~\bibnamefont {Devakul}},\
	}\bibfield  {title} {\emph {\bibinfo {title} {{How pairing mechanism dictates
					topology in valley-polarized superconductors with Berry curvature}},\ }}\href
	{https://doi.org/10.48550/arXiv.2503.05697} {\ }\Eprint
	{http://arxiv.org/abs/arXiv:2503.05697}{arXiv:2503.05697}\BibitemShut
	{NoStop}%
	\bibitem [{\citenamefont {Gaggioli}\ \emph {et~al.}()\citenamefont {Gaggioli},
		\citenamefont {Guerci},\ and\ \citenamefont {Fu}}]{Gaggioli_pentaG_2025}%
	\BibitemOpen
	\bibfield  {author} {\bibinfo {author} {\bibfnamefont {F.}~\bibnamefont
			{Gaggioli}}, \bibinfo {author} {\bibfnamefont {D.}~\bibnamefont {Guerci}}, \
		and\ \bibinfo {author} {\bibfnamefont {L.}~\bibnamefont {Fu}},\ }\bibfield
	{title} {\emph {\bibinfo {title} {{Spontaneous vortex-antivortex lattice and
					Majorana fermions in rhombohedral graphene}},\ }}\href
	{https://doi.org/10.48550/arXiv.2503.16384} {\ }\Eprint
	{http://arxiv.org/abs/arXiv:2503.16384}{arXiv:2503.16384}\BibitemShut
	{NoStop}%
	\bibitem [{\citenamefont {Dong}\ and\ \citenamefont {Lee}()}]{Dong2025}%
	\BibitemOpen
	\bibfield  {author} {\bibinfo {author} {\bibfnamefont {Z.}~\bibnamefont
			{Dong}}\ and\ \bibinfo {author} {\bibfnamefont {P.~A.}\ \bibnamefont {Lee}},\
	}\bibfield  {title} {\emph {\bibinfo {title} {A controllable theory of
				superconductivity due to strong repulsion in a polarized band},\ }}\href
	{https://arxiv.org/abs/2503.11079} {\ }\Eprint
	{http://arxiv.org/abs/arXiv:2503.11079}{arXiv:2503.11079}\BibitemShut
	{NoStop}%
	\bibitem [{\citenamefont {Christos}\ \emph
		{et~al.}({\natexlab{b}})\citenamefont {Christos}, \citenamefont {Bonetti},\
		and\ \citenamefont {Scheurer}}]{Christos_pentaG_2025}%
	\BibitemOpen
	\bibfield  {author} {\bibinfo {author} {\bibfnamefont {M.}~\bibnamefont
			{Christos}}, \bibinfo {author} {\bibfnamefont {P.~M.}\ \bibnamefont
			{Bonetti}}, \ and\ \bibinfo {author} {\bibfnamefont {M.~S.}\ \bibnamefont
			{Scheurer}},\ }\bibfield  {title} {\emph {\bibinfo {title} {{Finite-momentum
					pairing and superlattice superconductivity in valley-imbalanced rhombohedral
					graphene}},\ }}\href {https://doi.org/10.48550/arXiv.2503.15471} {\
		({\natexlab{b}})},\ \Eprint
	{http://arxiv.org/abs/arXiv:2503.15471}{arXiv:2503.15471}\BibitemShut
	{NoStop}%
	\bibitem [{\citenamefont {Chen}\ and\ \citenamefont
		{Schrade}()}]{Chen_pentaG_2025}%
	\BibitemOpen
	\bibfield  {author} {\bibinfo {author} {\bibfnamefont {Y.}~\bibnamefont
			{Chen}}\ and\ \bibinfo {author} {\bibfnamefont {C.}~\bibnamefont {Schrade}},\
	}\bibfield  {title} {\emph {\bibinfo {title} {{Intrinsic superconducting
					diode effect and nonreciprocal superconductivity in rhombohedral graphene
					multilayers}},\ }}\href {https://doi.org/10.48550/arXiv.2503.16391} {\
	}\Eprint
	{http://arxiv.org/abs/arXiv:2503.16391}{arXiv:2503.16391}\BibitemShut
	{NoStop}%
	\bibitem [{\citenamefont {Sedov}\ and\ \citenamefont
		{Scheurer}()}]{Sedov_probeFFLO_2025}%
	\BibitemOpen
	\bibfield  {author} {\bibinfo {author} {\bibfnamefont {D.}~\bibnamefont
			{Sedov}}\ and\ \bibinfo {author} {\bibfnamefont {M.~S.}\ \bibnamefont
			{Scheurer}},\ }\bibfield  {title} {\emph {\bibinfo {title} {{Probing
					superconductivity with tunneling spectroscopy in rhombohedral graphene}},\
	}}\href {https://doi.org/10.48550/arXiv.2503.12650} {\ }\Eprint
	{http://arxiv.org/abs/arXiv:2503.12650}{arXiv:2503.12650}\BibitemShut
	{NoStop}%
\end{thebibliography}

%

\newpage
\newpage
\setcounter{equation}{0}
\setcounter{figure}{0}
\setcounter{table}{0}
\setcounter{page}{1}
\makeatletter
\renewcommand{\theequation}{S\arabic{equation}}
\renewcommand{\thefigure}{S\arabic{figure}}
\renewcommand{\thetable}{S\arabic{table}}
\onecolumngrid

\section{Supplementary Material}
\subsection{A. Small-field expansions and Fermi surface deformations}\label{Sec:suppA}
\begin{figure}[!b]
\centering
\includegraphics[width=1.0\columnwidth]{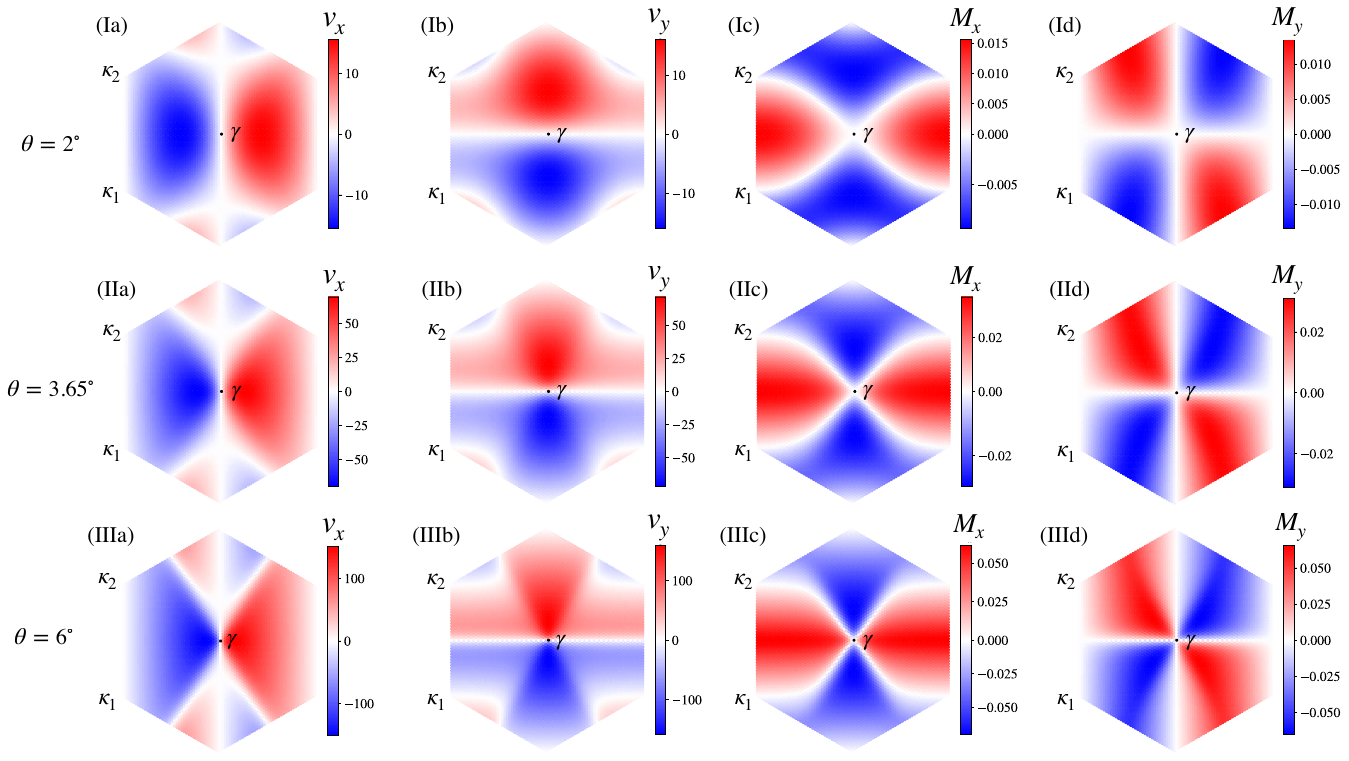}
\caption{\label{figS1_Mv} {
$\pmb{v}(\pmb{k}) \equiv \partial_{\pmb{k}} \varepsilon|_{B=0}$ and $\pmb{M}(\pmb{k}) \equiv  \partial_{\pmb{B}} \varepsilon|_{B=0}$ of valley $K$ for tWSe$_2$ with $V_z=0$. (Ia-Id) $\theta=2^\circ$, (IIa-IId) $3.65^\circ$ and (IIIa-IIId) $6^\circ$.
\pmb{v} is in unit of meV$\cdot$nm, and $\pmb{M}$ is in unit of meV/T.}
}
\end{figure}

\begin{figure}
\centering
\includegraphics[width=1.0\columnwidth]{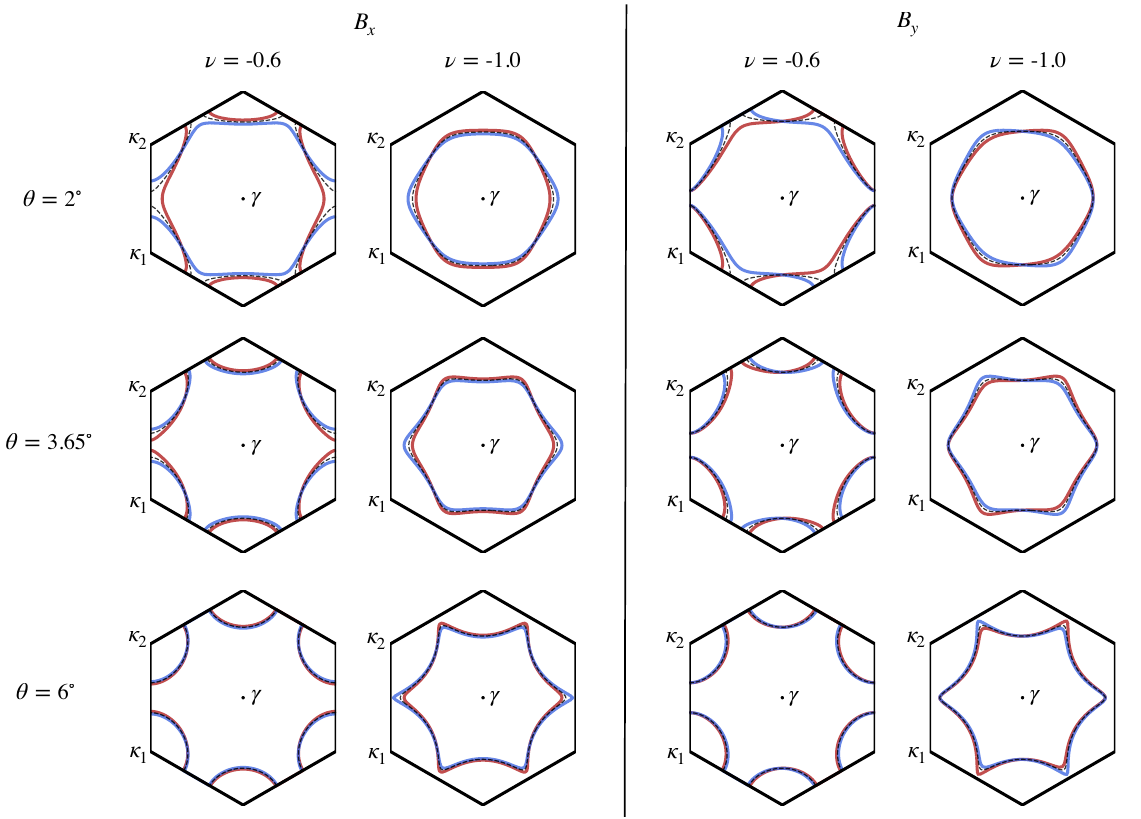}
\caption{\label{figS2_FS} {
Fermi surface contours of tWSe$_2$ at $V_z=0$ for $\theta=2^\circ$, $3.65^\circ$ and $6^\circ$.
Black dashed lines show the Fermi surface at $B=0$, which is identical for $\pm K$ valleys. 
Colored solid lines show Fermi surfaces under an in-plane field: $\pmb{B} = B_x \hat{x}$ (left) and $\pmb{B} = B_y \hat{y}$ (right), with red for valley $K$ and blue for valley $-K$. The Fermi surfaces are plotted in the MBZ of valley $K$, where the blue contours are obtained by applying 2D inversion to those of valley $-K$.
To highlight the Fermi surface distortions, large magnetic fields are used: $B = 10$ T for $\theta=2^\circ$ and $B = 20$ T for $\theta=3.65^\circ$, $6^\circ$.
}}
\end{figure}

To streamline numerical calculations, we expand $\varepsilon(\pmb{k},\pmb{B})$ and $u_{\pmb{k},l}(\pmb{q}, \pmb{B})$ in Eq.~(\ref{Eq_chillp}) in terms of small momentum and weak field:
\begin{align}
u_{\pmb{k},l}(\pmb{q}, \pmb{B}) \approx u_{\pmb{k},l}(q=0, B=0) = \sum\limits_{\pmb{G}} |z_{+,l,\pmb{G}}(\pmb{k})|^2, \\
\varepsilon(\pmb{k}_0 + \pmb{q}, \pmb{B}) \approx \varepsilon(\pmb{k}_0,B=0) + \pmb{v}(\pmb{k}_0) \cdot \pmb{q} + \pmb{M}(\pmb{k}_0) \cdot \pmb{B}, \\
\end{align}
where
\begin{equation}
\begin{split}
\pmb{v}(\pmb{k}_0) &= \partial_{\pmb{k}} \varepsilon|_{\pmb{k}=\pmb{k}_0, B=0} =\langle\partial_{\pmb{k}}H(\pmb{k},\pmb{B}) \rangle|_{\pmb{k}=\pmb{k}_0, B=0}, \\
\pmb{M}(\pmb{k}_0) &= \partial_{\pmb{B}} \varepsilon|_{\pmb{k}=\pmb{k}_0, B=0}
= \langle\partial_{\pmb{B}}H(\pmb{k},\pmb{B}) \rangle|_{\pmb{k}=\pmb{k}_0, B=0}.
\end{split}
\end{equation}
The $k$-space distributions of $\pmb{v}(\pmb{k})$ and $\pmb{M}(\pmb{k})$ for twist angles $\theta=2^\circ, 3.65^\circ$ and $6^\circ$ are shown in Fig.~\ref{figS1_Mv}. 
As $\theta$ increases, the effective interlayer coupling weakens, both $\pmb{v}$ and $\pmb{M}$ increase (Fig.~\ref{figS1_Mv}), but their ratio $M/v$ decreases. As a result, the overall $q/q_B$ of the FF phase decreases with increasing $\theta$ (Figs.~\ref{figS3}-\ref{figS4}).
For smaller twist angles, where the moir\'e bands are flatter, an in-plane field has relatively stronger orbital effects, distorting the Fermi surface more significantly. This trend is reflected by the overall ratio $M/v$ and Fermi surface deformations at different twist angles (Fig.~\ref{figS2_FS}). As $\theta$ increases, the FF phase extends over a broader region in the $(B,T)$ phase diagram, with the tricritical point at $V_z=0$ moving to higher temperatures and smaller $B$ fields (Figs.~\ref{figS3}-\ref{figS4}). This suggests that a larger twist angle requires a smaller in-plane field to induce the FF state.
In addition, the FF phase is often suppressed near VHS, as seen in Fig.~\ref{figS3} (Ib, IIb) and Fig.~\ref{figS4} (Ic), where the in-plane field significantly distorts the Fermi surfaces of opposite valleys into completely different topologies, preventing effective pairing.

At $V_z=0$, the field-induced in-plane orbital magnetizations satisfy $M_x \propto - v_y \odot l_z$ and $M_y \propto v_x \odot l_z$, where $\odot$ denotes element-wise matrix multiplication. This implies that $M_x$ is symmetric under both $k_x \rightarrow -k_x$ and $k_y \rightarrow -k_y$ in the valley-projected MBZ, whereas $M_y$ is antisymmetric (see Fig.~\ref{figS1_Mv}). As a result, under $B_y$, the FF phase is less favorable when the Fermi surface contour crosses both $k_x=0$ and $k_y=0$, as illustrated in Fig.~\ref{figS5}(b).
This can be partially understood by considering the Fermi surface deformations shown in Fig.~\ref{figS2_FS}. If every point on a Fermi surface contour has a ``nesting'' partner requiring a different $\pmb{q}$, no single $\pmb{q}$ is preferred and FF pairing is suppressed. This scenario is more likely under $B_y$ than under $B_x$, as evident from Figs.~\ref{figS1_Mv}-\ref{figS2_FS}.
In the opposite limit, where every Fermi surface point pairs with a partner at the same $\pmb{q}$, the upper critical in-plane field can, in principle, be arbitrarily large (ignoring Zeeman effects). This is the case when the two layers are completely decoupled.
The dependence on field direction becomes more complex when the detailed distributions of $\pmb{v}(\pmb{k})$ and $\pmb{M}(\pmb{k})$, Fermi surface geometry, and layer polarization are considered, especially for $V_z \neq 0$.

\subsection{B. Twist-angle dependence of phase diagrams under $B_x$}
\begin{figure}
\centering
\includegraphics[width=1.0\columnwidth]{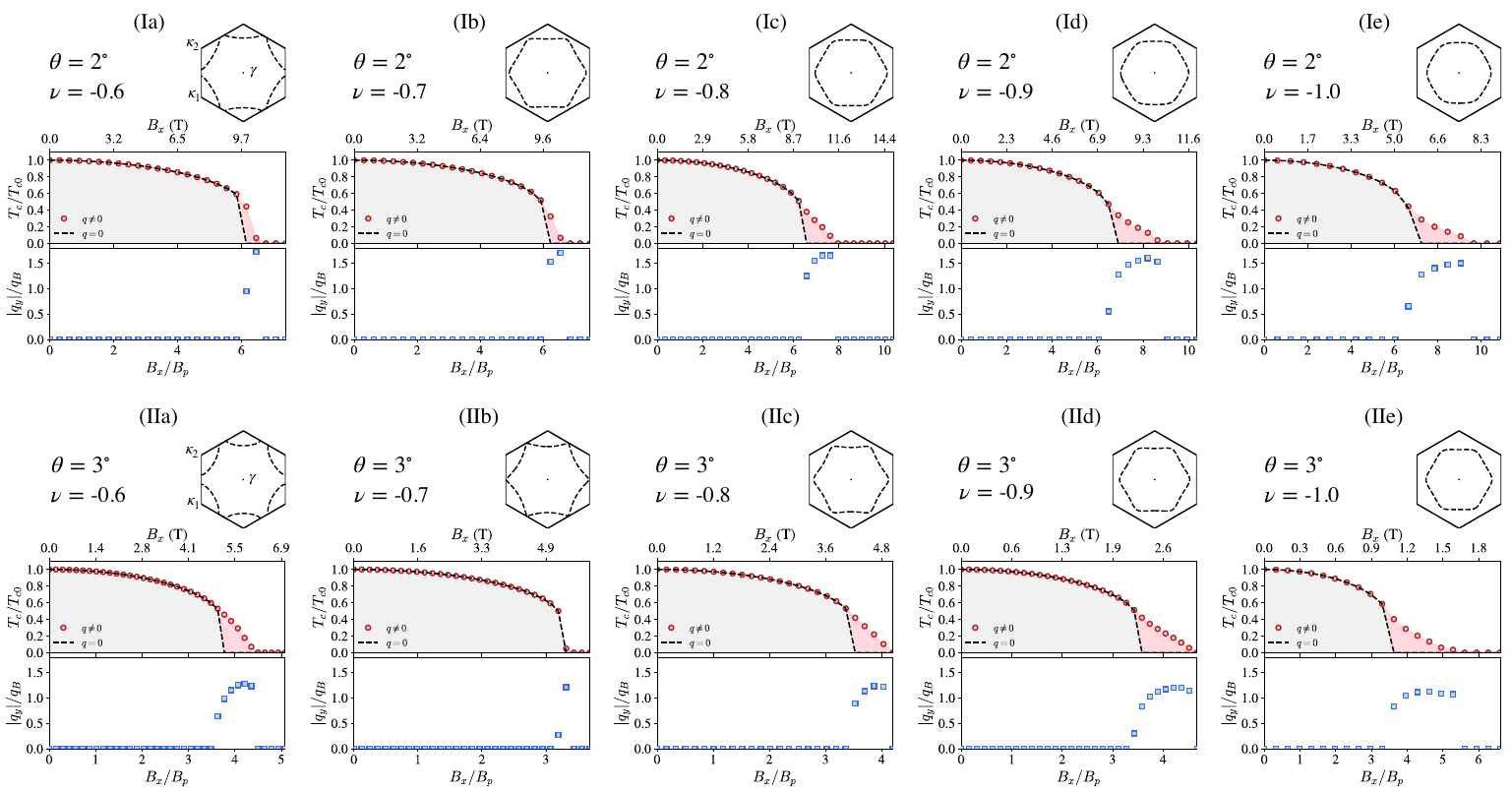}
\caption{\label{figS3} {
$T_c$ vs. $B_x$ phase diagram and the Cooper pair momentum corresponding to the highest $T_c$ for tWSe$_2$ at $V_z=0$ with twist angles: (Ia-Ie) $\theta=2^\circ$ and (IIa-IIe) $\theta=3^\circ$. The black dashed lines in the hexagonal insets represent Fermi surface contours for the respective filling factors indicated in each subfigure.
}}
\end{figure}

\begin{figure}
\centering
\includegraphics[width=1.0\columnwidth]{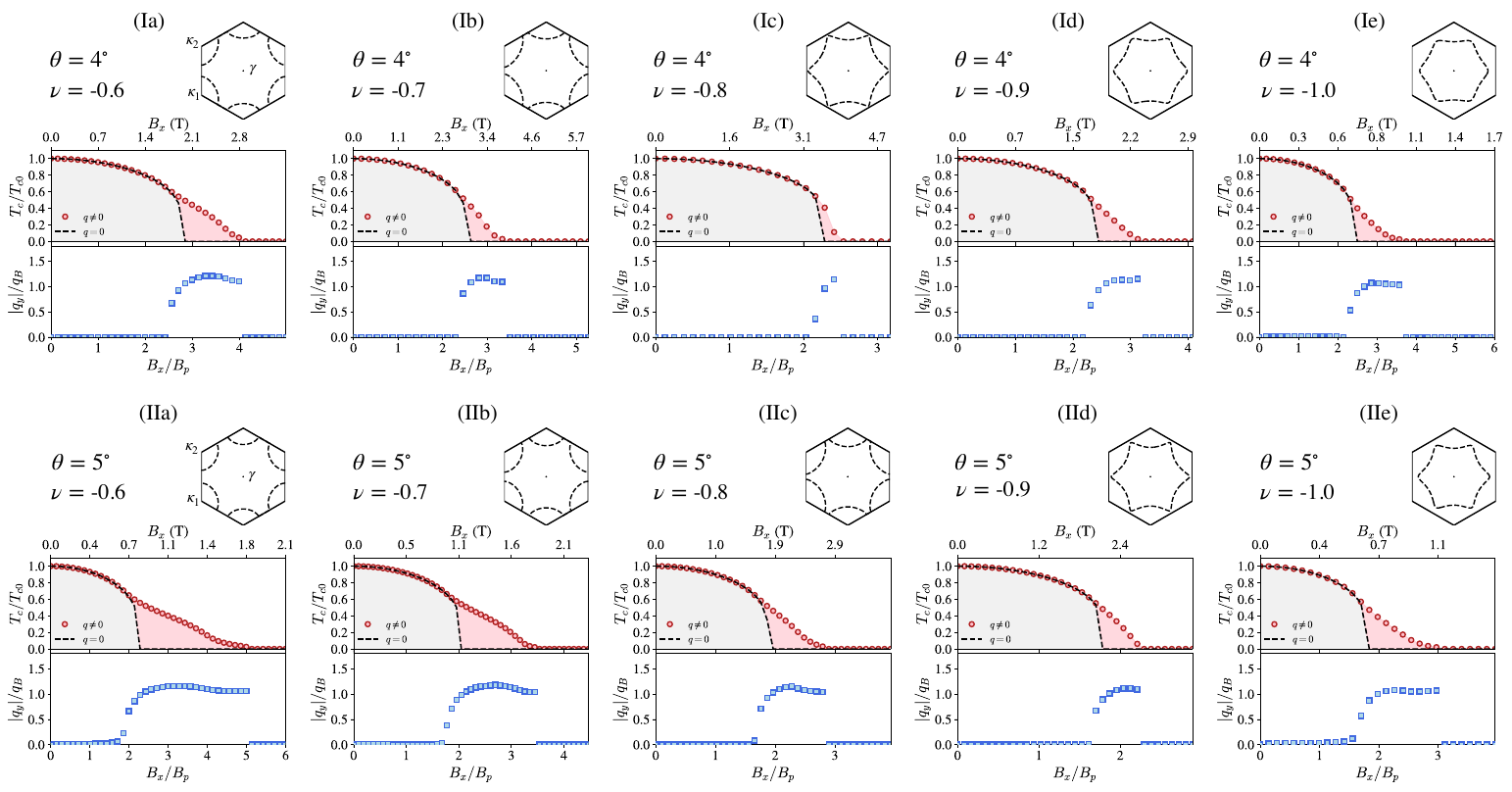}
\caption{\label{figS4} {
Same as Fig.~\ref{figS3}, but for twist angles: (Ia-Ie) $\theta=4^\circ$ and (IIa-IIe) $\theta=5^\circ$.
}}
\end{figure}

In Figs.~\ref{figS3}-\ref{figS4}, we explore $T_c$ vs. $B_x$ phase diagrams for twist angles ranging from $2^\circ$-$5^\circ$ at $V_z=0$. As discussed in Sec.~\ref{Sec:suppA}, increasing $\theta$ reduces the overall $q/q_B$ of the FF phase and expands the FF region in the $(B,T)$ phase diagram. The tricritical point at $V_z=0$ shifts to higher temperatures and lower $B$ fields with increasing $\theta$ (Figs.~\ref{figS3}-\ref{figS4}), indicating that a larger twist angle requires a weaker in-plane field to induce the FF state.
Near the VHS, the FF phase is suppressed, as seen in Fig.~\ref{figS3} (Ib, IIb) and Fig.~\ref{figS4} (Ic).

\subsection{C. Phase diagrams under $B_y$}
\begin{figure}
\centering
\includegraphics[width=1.0\columnwidth]{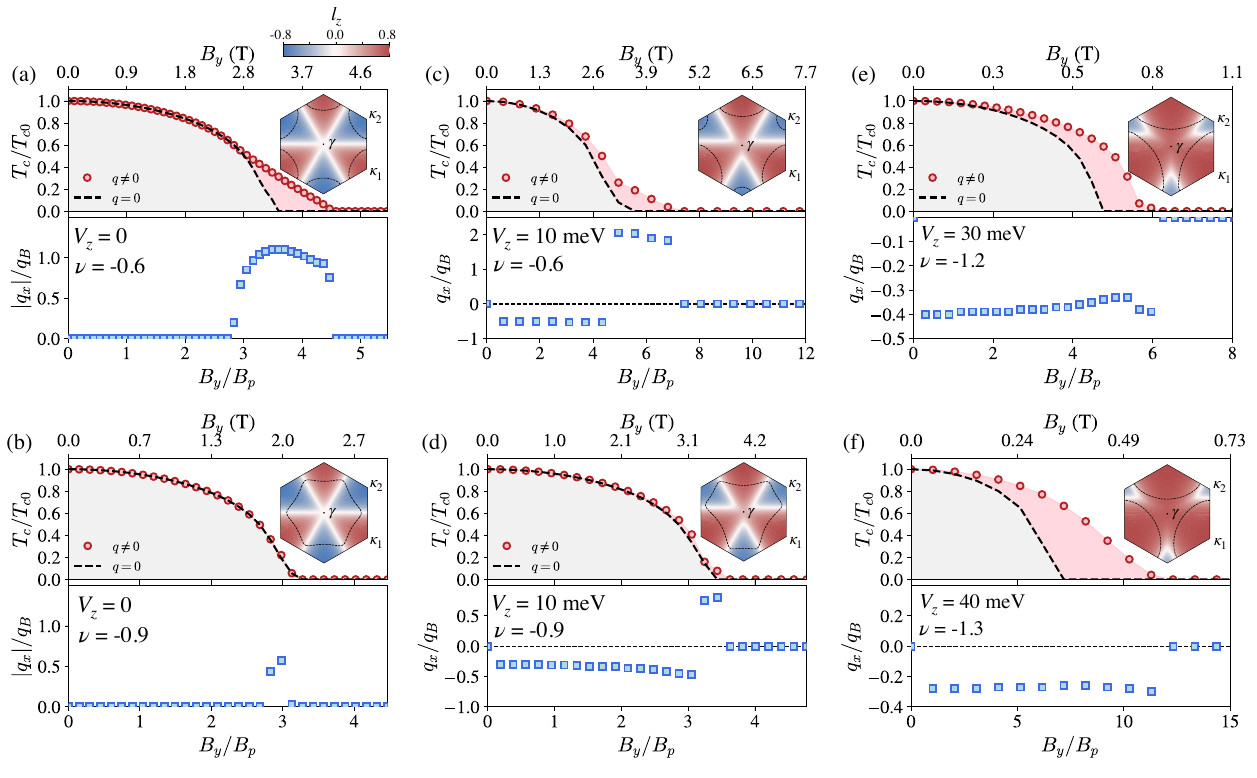}
\caption{\label{figS5} {
Phase diagrams under $\pmb{B} = B_y\hat{y}$, with the same $V_z$ and filling factors as in Figs.~2,3 of the main text, as indicated in each subfigure.
}}
\end{figure}

Figure~\ref{figS5} presents the $T_c$ vs. $B_y$ phase diagrams under the same $V_z$ and $\nu$ conditions as those considered for $B_x$ in the main text, allowing a direct comparison.
As discussed in Sec.~\ref{Sec:suppA},
at $V_z=0$ and under $B_y$, when the Fermi surface contour crosses both $k_x=0$ and $k_y=0$,
the FF phase is less favorable (Fig.~\ref{figS5}(b)). A similar suppression occurs in Fig.~\ref{figS5}(d) at small $V_z$. In other cases shown in Fig.~\ref{figS5}, the FF phase exhibits general properties consistent with those in Figs.~\ref{fig_TcvsB_Vz0}-\ref{fig_TcvsB_Vz30to60} in the main text: A phase transition within the FF phase appears when layer symmetry is broken, and two separate Fermi pockets form at the two moir\'e mini-valleys (Fig.~\ref{figS5}(c)); As $|V_z|$ increases, the FF phase region expands (Fig.~\ref{figS5}(e,f)). Note that an extra phase transition occurs in Fig.~\ref{figS5}(d), though its phase space region is extremely small, which is attributed to the detailed properties of $\pmb{v}(\pmb{k})$, $\pmb{M}(\pmb{k})$ and Fermi surface.

\end{document}